\documentclass[conference]{IEEEtran}
\IEEEoverridecommandlockouts
\def\ps@headings{%
\def\@oddhead{\mbox{}\scriptsize\rightmark \hfil \thepage}%
\def\@evenhead{\scriptsize\thepage \hfil \leftmark\mbox{}}%
\def\@oddfoot{}%
\def\@evenfoot{}}
\usepackage{epsfig, epsf, amsmath, array, amsthm, amssymb, latexsym, graphics, multirow}
\usepackage{graphicx}
\usepackage{amsthm}
\usepackage{bm}
\usepackage{amsmath}
\newcommand {\mymarginpar}[1]{\marginpar{#1}}
\renewcommand {\marginpar}[1]{} % comment out this command to show labels in the margin

\def\_{\rule{.3em}{.15ex}}      % Get underscore by typing \_.

\newcommand{\ls}[1]
   {\dimen0=\fontdimen6\the\font
    \lineskip=#1\dimen0
    \advance\lineskip.5\fontdimen5\the\font
    \advance\lineskip-\dimen0
    \lineskiplimit=.9\lineskip
    \baselineskip=\lineskip
    \advance\baselineskip\dimen0
    \normallineskip\lineskip
    \normallineskiplimit\lineskiplimit
    \normalbaselineskip\baselineskip
    \ignorespaces
   }
\newcommand {\bearn}{\begin{eqnarray*}}
\newcommand {\eearn}{\end{eqnarray*}}
\newcommand {\barr}{\begin{array}}
\newcommand {\earr}{\end{array}}

\newcommand {\N}{{\cal N}}

\def\defeq{\stackrel{\scriptstyle\rm def}{=}}

\newtheorem{definition}{Definition}
\newtheorem{property}[definition]{Property}
\newtheorem{proposition}[definition]{Proposition}
\newtheorem{lemma}[definition]{Lemma}
\newtheorem{theorem}[definition]{Theorem}
\newtheorem{corollary}[definition]{Corollary}
\newtheorem{example}[definition]{Example}
\newtheorem{remark}[definition]{Remark}
\newtheorem{algorithm}[definition]{Algorithm}
\newtheorem{assumption}[definition]{Assumption}

\newcommand {\benum} {\begin{enumerate}}
\newcommand {\eenum} {\end{enumerate}}

\newcommand {\bdesc} {\begin{description}}
\newcommand {\edesc} {\end{description}}

\newcommand {\bfig}[2] {\begin{figure}[htbp]
                        \centerline {
                         \epsfig{figure={#1},clip=,width={#2}}}}
\newcommand {\brotatefig}[2] {\begin{figure}[htbp]
                        \centerline {
                         \epsfig{figure={#1},clip=,angle=-90,width={#2}}}}
\newcommand {\bfigfirst}[2] {\begin{figure}[h]
                        \centerline {
                        \setlength{\epsfxsize}{#2}
                        \epsffile{#1}}}
\newcommand {\efig}[2]{ \caption{#2}
                        \label{fig:#1}
                        \end{figure}
                        \centerline {
                        \mymarginpar{fig:#1}}}
\newcommand {\erotatefig}[2]{ \caption{#2}
                        \label{fig:#1}
                        \end{figure}
                        \mymarginpar{fig:#1}}
\newcommand {\rfig}[1]{Figure \ref{fig:#1}}

\newcommand {\btab}[1]{
                       \begin{table}
                       \centering
                       \begin{tabular}{#1}}
\newcommand {\etab}[3] {
                       \end{tabular}
                       \caption[#3]{#2}
                       \label{tab:#1}
                       \end{table}
                       \mymarginpar{tab:#1}
                       \vspace{.1in}}

\newcommand {\btabular}[1]{\begin{center}
                       \begin{tabular}{#1}}
\newcommand {\etabular}{\end{tabular}
                       \end{center}}

\newcommand {\bdefin}[1]{\begin{definition}
                      \mymarginpar{def:#1}
                      \label{def:#1} }
\newcommand {\edefin}       {\end{definition}}

\newcommand {\bassum}[1]{\begin{assumption}
                      \mymarginpar{ass:#1}
                      \label{ass:#1} }
\newcommand {\eassum}       {\end{assumption}}

\newcommand {\bpro}[1]{\begin{property}
                      \mymarginpar{pro:#1}
                      \label{pro:#1} }
\newcommand {\epro}   {\end{property}}

\newcommand {\bprop}[1]{\begin{proposition}
                      \mymarginpar{prop:#1}
                      \label{prop:#1} }
\newcommand {\eprop}       {\end{proposition}}
\newcommand {\rprop}[1]{Proposition \ref{prop:#1}}

\newcommand {\blem}[1]{\begin{lemma}
                      \mymarginpar{lem:#1}
                      \label{lem:#1} }
\newcommand {\elem}   {\end{lemma}}

\newcommand {\bthe}[1]{\begin{theorem}
                      \mymarginpar{the:#1}
                      \label{the:#1} }
\newcommand {\ethe}   {\end{theorem}}

\newcommand {\bcor}[1]{\begin{corollary}
                      \mymarginpar{cor:#1}
                      \label{cor:#1} }
\newcommand {\ecor}   {\end{corollary}}

\newcommand {\bax}[1]{\begin{axiom}
                      \mymarginpar{ax:#1}
                      \label{ax:#1} }
\newcommand {\eax}       {\vspace{-.1in} \end{axiom}}

\newcommand {\bex}[2]{\vspace{.1in}
                      \begin{example}
                      \mymarginpar{ex:#1}
                       {\bf #2}
                      \label{ex:#1} \em}
\newcommand {\eex}       {\end{example} \vspace{.3cm} }

\newcommand {\brem}[1]{\begin{remark}
                      \mymarginpar{rem:#1}
                      \label{rem:#1} \em }
\newcommand {\erem}   {\end{remark}}

\newcommand {\beq}[1]{\mymarginpar{eq:#1}
                      \begin{equation}
                      \label{eq:#1} }

\newcommand {\beqno}[1]{\mymarginpar{eq:#1}
                      \begin{eqnarray}
                      \nonumber}

\newcommand {\eeq}       {\end{equation}}
\newcommand {\eeqno}       { && \end{eqnarray}}
\newcommand {\req}[1]{(\ref{eq:#1})}

\newcommand {\bear}[1]{\mymarginpar{eq:#1}
                       \begin{eqnarray}
                       \label{eq:#1} }

\newcommand {\bearno}[1]{\mymarginpar{eq:#1}
                       \begin{eqnarray}
                       \nonumber}

\newcommand {\eear}{\end{eqnarray}}
\newcommand {\eearno}{\end{eqnarray}}
\newcommand {\bsel}{\left \{ \begin{array}{cl}}
\newcommand {\esel}{\end{array} \right.}

\newcommand {\bmat}[1]{\left [ \begin{array}{#1}}
\newcommand {\emat}{\end{array} \right ]}
\def\R{I\kern-0.30em R}
\def\N{I\kern-0.30em N}
\def\P{I\kern-0.30em P}

\def\ex{{\bf\sf E}}
\def\pr{{\bf\sf P}}

\def\bfa{{\mbox{$\bm{a}$}}}

\def\bff{{\mbox{$\bm{f}$}}}

\def\bfu{{\mbox{$\bm{u}$}}}

\def\bfx{{\mbox{$\bm{x}$}}}

\def\bfsx{{\mbox{\scriptsize$\bm{x}$}}}

\def\bfA{{\mbox{$\bm{A}$}}}

\def\bfI{{\mbox{$\bm{I}$}}}
\def\bfJ{{\mbox{$\bm{J}$}}}

\def\bfX{{\mbox{$\bm{X}$}}}

\def\bfone{{\mbox{$\bm{1}$}}}

\def\argmax{\mathop{\rm argmax}}
\def\indicatorFunction#1{{1_{\left\{#1\right\}}}}

\newcommand{\binomcoeff}[2]{\left(\begin{array}{c}#1\\#2\end{array}\right)}

\usepackage{abstract}
\usepackage{amsmath}
\usepackage{multicol}
\usepackage{graphicx}
\usepackage{indentfirst}
\usepackage{amsfonts}
\usepackage{color}
\usepackage{cite}
\usepackage{lipsum}
\usepackage{mathtools}
\usepackage{cuted}
\usepackage{url}
\usepackage[ruled]{algorithm}
\usepackage{algorithmic}

\title{Epidemic Spreading in a Social Network with Facial Masks wearing Individuals}
\author{Duan-Shin Lee,~\IEEEmembership{Senior Member,~IEEE,}
	and Miao Zhu\thanks{The authors are with the Institute of Communications
		Engineering, National Tsing Hua University, Hsinchu 30013, Taiwan, R.O.C.
		Email: lds@cs.nthu.edu.tw, % cschang@ee.nthu.edu.tw,
		xmzhumiao11111@hotmail.com.
		This research was supported in part by the Ministry of Science and Technology,
		Taiwan, R.O.C., under Contract 109-2221-E-007-093-MY2.
}}
\date{\today}
\begin{document}
\maketitle
\begin{abstract}
In this paper, we present a susceptible-infected-recovered (SIR) model with
individuals wearing facial masks and individuals who do not.
The disease transmission rates, the recovering rates and the fraction of
individuals who wear masks are all time dependent in the model.
We develop a progressive estimation of the disease transmission rates
and the recovering rates based on the COVID-19 data published by John Hopkins University.
We determine the fraction of individual who wear masks by a maximum
likelihood estimation, which maximizes the transition probability of
a stochastic susceptible-infected-recovered model.  The transition probability is
numerically difficult to compute if the number of infected individuals is large.
We develop an approximation for the transition probability based on
central limit theorem and mean field approximation.
We show through numerical study that our approximation works well.
We develop a bond percolation analysis to predict the eventual fraction
of population who are infected, assuming that parameters of the SIR model
do not change anymore.

We predict the outcome of COVID-19 pandemic using our theory.

\end{abstract}
\textbf{keywords}:bond percolation; epidemic network; susceptible-infected-recovered
model; masks; COVID-19

\section{Introduction}\label{SI}

In December of 2019 a few patients of a new infectious respiratory disease were detected
in Wuhan, China.  This disease has been called coronavirus disease 2019 (COVID-19)
and the virus that causes COVID-19 has been named SARS-CoV-2 by
the World Health Organization (WHO).  WHO declared the outbreak a public health emergency
of international concern at the end of January 2020, and a pandemic on March 11, 2020.
Since the outbreak, most countries have adopted various measures in an attempt to contain
the pandemic.  These measures include restriction of travelling, shutting down schools, restaurants
and businesses, cancelling large gatherings such as concerts, sports and religious activities,
and even city lockdowns where residents are not allowed to leave home unless emergencies.
Clearly these measures seriously affect daily lives and are devastating to
the economics.  The purpose of this paper is to show that wearing facial masks is a simple
and inexpensive measure to contain the spread of COVID-19.  In fact, we shall show that
if a relatively small fraction of population wears facial masks, the disease can be contained.

Facial masks have been shown in labs to be effective to limit the spread of droplets or aerosols if
a wearer coughs
\cite{Lai2012Effectiveness,Brienen2010,Davis2013Testing,Patel2016Respiratory}.
This ability is measured by a quantity called the outward mask filter efficiency of masks.
Facial masks also protect their wearers from inhaling droplets or aerosols from
a nearby cougher, if the cougher does not wear a
mask.  This ability is measured by a quantity called the inward mask filter efficiency
of masks. Thus, facial masks can be particularly useful to confine the
spreading of diseases that transmit through  droplets or aerosols.
However, the effect of wearing facial masks to the epidemic spreading
has never been studied in a network level.
In this paper we present an epidemic network study to justify this argument.
Specifically, we propose a time dependent susceptible-infected-recovered (SIR)
model with two types of individuals. Type 1 individuals wear a facial mask and type 2
individuals do not.  A randomly selected individual from a population is a type 1
individual with probability $p$, and is of type 2 with probability $1-p$.
There are four types of contacts between two individuals
depending on whether the two individuals wear a facial mask or not.
These four types of contacts have four different disease transmission rates.
From the data published by John Hopkins University \cite{John-Hopkins2020}
we progressively estimate the time dependent disease transmission rates
and the recovery rates of the SIR model.

For parameter $p$, we propose a stochastic version of the SIR model.
Specifically, we assume that the COVID-19 epidemic propagates in a social contact
network according to an independent cascade model \cite{Kempe2003, Easley2010}.
The social contact network in our study consists of tree connection of
cliques of random sizes.  We derive
the transition probability of the number of infected individuals from one time
slot to the next.
We propose a maximum likelihood estimation of $p$ that maximizes
the transition probability.  The transition probability is expressed in terms
of binomial distributions.  Parameters of the binomial distributions
corresponding to real data published
in \cite{John-Hopkins2020} are typically very large.  That makes the transition
probability numerically difficult to compute.  We propose an approximation
of the transition probability based on central limit theorem and
mean field approximation.  Through numerical studies, we show that the
approximation works well.
We derive a percolation analysis of the maximum number
of individuals that can eventually be infected.  We incorporate the maximum likelihood
estimation and the percolation analysis into the progressive estimation.  That is,
based on the data published by John Hopkins University, we progressively estimate
the disease transmission rates and the recovery rates.  We then find $p$ from the
maximum likelihood estimation.  Finally, using percolation analysis, we predict
the maximum number of individuals that can eventually be infected.

COVID-19 has attracted a lot of research work on epidemic networks.  We now review
some research work that is related to the use of facial masks.  Li et al. \cite{Li.2020} proposed
a decision making process in a susceptible-exposed-symptomatic-quarantined model.
At each time, every susceptible individual chooses whether to wear a mask or not in the next time
step, which depends on an evaluation of the potential costs and perceived risk of infection.
Damette \cite{Damette.2020} conducted a panel econometric exercise to assess the dynamic impact of
face mask use on both infected cases and fatalities at a global scale.
A negative impact of mask
wearing on fatality rates and on the COVID-19 number of infected cases was observed.
Damette found that population density and pollution levels
are significant determinants of heterogeneity regarding mask adoption across
countries, while altruism, trust in government and demographics are not.
In addition, the most effective way to increase mask use is through strict laws by governments.
Betsch \cite{Betch.2020} examined nearly 7000 German participants and
observed that implementing a mandatory policy increased actual compliance of wearing masks.
Mask wearing correlated positively with other protective behaviours.  They also observed
that voluntary policy is far less effective, and mandatory policy is far more effective to
curb transmission of airborne virus.

The outline of this paper is as follows.  In Section \ref{TSIR}, we present a
time dependent SIR model.  We present a progressive estimation of the disease
transmission rates and the recovery rates of the model.  In Section
\ref{SEP}, we present a social contact network consisting of tree connection
of cliques.  We present a maximum likelihood method to estimate $p$.
In Section \ref{CC}, we derive the clustering coefficient of this social contact network.
In Section \ref{SPA} we present a percolation
analysis.  In Section \ref{SNS} we present results of numerical study
and simulation.  We present the conclusions in Section \ref{SC}.

\section{Time Dependent SIR Model}\label{TSIR}

In this section we present a discrete-time susceptible-infected-recovered (SIR)
model.  SIR models have long been used to model and study
epidemics \cite{Newman2010}.  We now survey a few recent research studies that
are related to SIR models.  Opuszko et al. \cite{Opuszko.2013} studied the impact
of the network structure on the SIR model spreading phenomena.
This study is based on simulation of SIR models on real life online social networks
as well as mathematical network models.
Bernardes et al. \cite{Bernardes.2012} studied SIR models on P2P systems.
The study was also simulation based.
Wang et al. \cite{Wang.2019} and Zheng et al.
\cite{Zheng.2018} used two-layer multiplex networks
to investigate the multiple influence between awareness diffusion and epidemic
propagation, where the upper layer represents the awareness diffusion regarding
epidemics and the lower layer expresses the epidemic propagation.
Wang et al. \cite{Wang.2019} analytically showed that the epidemic threshold
is correlated with the  awareness diffusion as well as the topology of epidemic networks.

In our model time is divided into periods of equal lengths.  There are two
types of individuals.  Type 1 individuals wear a facial mask and type 2
individuals do not.  A randomly selected individual from a population is a type 1
individual with probability $p(t)$ in period $t$.  A randomly selected individual is of type 2
with probability $1-p(t)$.  Let $s(t)$ and $r(t)$ be the number of susceptible and recovered
individuals, respectively, at time $t$.  Similarly, let $x_i(t)$ be
the number of infected type $i$ individuals in period $t$ for $i=1, 2$.
In this model, the disease transmission rates
are time dependent.  Let $\beta_{ij}(t)$ be the expected number
of type $j$ susceptible individuals who receive the disease from one type $i$ infected
individual per unit time in period $t$.  Let $\gamma(t)$ be the recovering rate of
the disease in period $t$.  We assume that both types of infected individuals have
the same recovering rate.

\iffalse
Let $\gamma(t)$ be the recovering rate of the disease in period $t$.
According to European Centre for Disease Prevention and Control \cite{EU-CDC2020},
the RNA of the SARS-CoV-2 virus can be detected 1-2 days before the onset of symptoms
and it can persist for up to eight days in mild cases, and for longer periods in more severe cases.
Prolonged viral RNA shedding for up to 67 days has been reported \cite{Perera2020}.
In our study, we assume that on average COVID-19 takes 14 days to recover.
This is also the number of days that many countries quarantine their international
visitors.
\fi

In this paper we assume that the length of a time unit in the discrete-time
SIR model is $\tau$ days.
The dynamics of this discrete-time SIR model is as follows.  The existing infected
individuals at time $t-$ transmit the disease to newly infected individuals.
Those infected individuals existed at time $t-$ become recovered at time $(t+1)-$.
It is easy to derive
the following set of difference equations for the SIR model.
\begin{align}
&s(t+1)-s(t)\nonumber\\
&\quad=-\left(\sum_{i=1}^2 x_i(t)[\beta_{i,1}(t)p(t)+\beta_{i,2}(t)(1-p(t))]\right)
\frac{s(t)}{n}\nonumber\\
&x_1(t+1)-x_1(t)=x_1(t)\beta_{11}(t)\frac{s(t)p(t)}{n}\nonumber\\
&\quad +x_2(t)\beta_{21}(t)\frac{s(t)p(t)}{n}-\gamma(t) x_1(t)\nonumber \\
&x_2(t+1)-x_2(t)=x_1(t)\beta_{12}(t)\frac{s(t)(1-p(t))}{n}\nonumber\\
&\quad +x_2(t)\beta_{22}(t)\frac{s(t)(1-p(t))}{n}
-\gamma(t) x_2(t)\nonumber\\
&r(t+1)-r(t)= \gamma(t)(x_1(t)+x_2(t)),\label{update-r}
\end{align}
where
\[
n=s(t) +x_1(t)+x_2(t)+r(t)
\]
is the size of the population.  We assume that the epidemics is in the early stage.
That is, we assume that $s(t)\approx n$.  Under this assumption, the difference equations
for $x_1(t)$ and $x_2(t)$ reduce to
\begin{align}
x_1(t+1)-x_1(t)
&= (x_1(t)\beta_{11}(t)+x_2(t)\beta_{21}(t))p(t)\nonumber\\
&\quad -\gamma(t) x_1(t) \label{x1}\\
x_2(t+1) -x_2(t)
&= (x_1(t)\beta_{12}(t)+x_2(t)\beta_{22}(t))(1-p(t))\nonumber\\
&\quad-\gamma(t) x_2(t).\label{x2}
\end{align}

We now determine parameters $\beta_{ij}(t)$ and $p(t)$.
We reduce the number of parameters.
Previous study \cite{Brienen2010} suggested that
\begin{align}
\beta_{12}(t)&=(1-\eta_1)\beta_{22}(t) \label{beta12-assumption}\\
\beta_{21}(t)&=(1-\eta_2)\beta_{22}(t),\label{beta21-assumption}
\end{align}
where $\eta_1$ and $\eta_2$ are outward and inward efficiencies of masks, respectively.
Study \cite{Patel2016Respiratory} suggested that $\eta_1  > \eta_2$.
In addition, we assume that
\begin{equation}\label{beta11-assumption}
\beta_{11}(t)=(1-\eta_1)(1-\eta_2)\beta_{22}(t).
\end{equation}
Thus, we only need to determine $\beta_{22}(t)$.  The other three parameters
are determined according to Eqs. (\ref{beta12-assumption}), (\ref{beta21-assumption})
and (\ref{beta11-assumption}).  Substituting (\ref{beta12-assumption}), (\ref{beta21-assumption})
and (\ref{beta11-assumption}) into (\ref{x1}) and (\ref{x2}), we obtain
\begin{align}
x_1(t+1)
&= \beta_{22}(t)[x_1(t)(1-\eta_1)(1-\eta_2)\nonumber\\
&\quad+x_2(t)(1-\eta_2)]p(t)+(1-\gamma(t))x_1(t)\label{xx1}\\
x_2(t+1)
&= \beta_{22}(t)[x_1(t)(1-\eta_1)+x_2(t)](1-p(t))\nonumber\\
&\quad +(1-\gamma(t))x_2(t).\label{xx2}
\end{align}

Now we determine the values of parameters $\gamma(t)$, $\beta_{22}(t)$
and $p(t)$ based on the data published by John Hopkins University \cite{John-Hopkins2020}.
Note that John Hopkins University publishes total number
of daily newly infected individuals and the number of recovered individuals.
From the published data, one can easily compute $r(t)$ for each $t$.
The website does not distinguish between
infected individuals who wear masks or who do not.  Thus, we must determine
$\beta_{22}(t)$ and $p(t)$ based on the total number of infected individuals
\[
x(t)=x_1(t)+x_2(t)
\]
at time $t$.
From the data published by John Hopkins University, we estimate $\gamma(t)$,
$\beta_{22}(t)$ and $p(t)$.  We develop in section \ref{SPA} a percolation
analysis to determine the ultimate size of infected population,
if these parameters do not change.

Recovering rate is simple to determine.  From (\ref{update-r}) we have
\begin{align}
\gamma(t)=\frac{r(t+1)-r(t)}{x(t)}.\label{gamma-est}
\end{align}
Next, we determine $\beta_{22}(t)$.
To determine the value of $\beta_{22}(t)$ in period $t$, we let
$x(t)$ be the total number of infected individuals in period $t$, i.e.

Adding Eqs. (\ref{xx1}) and (\ref{xx2}) and solving for $\beta_{22}(t)$, we obtain
(\ref{beta22-approx}) displayed in Fig. \ref{fig-beta22-approx}.

%\begin{strip}
\begin{figure*}[!htbp]
\begin{align}
\beta_{22}(t)=\frac{x(t+1)-(1-\gamma(t))x(t)}{x_1(t)[(1-\eta_1)(1-\eta_2)p(t)
	+(1-\eta_1)(1-p(t))]+
x_2(t)[(1-\eta_2)p(t)+(1-p(t))]}.\label{beta22-approx}
\end{align}
\caption{An expression of $\beta_{22}(t)$.}\label{fig-beta22-approx}
\end{figure*}
%\end{strip}
To determine the value of $\beta_{22}(t)$ at time $t$, we use $x(t+1)$ published in
\cite{John-Hopkins2020}.  We assume that $x_1(t)$ and $x_2(t)$ and $p(t)$ are
available for $t$.  We use (\ref{beta22-approx}) to determine $\beta_{22}(t)$.
The value of $p(t)$ is determined by a maximum
likelihood estimation method.  We shall present this method in Section \ref{SEP}.
After $\beta_{22}(t)$ at time $t$ is determined, we use (\ref{xx1})
and (\ref{xx2}) to determine $x_1(t+1)$ and $x_2(t+1)$.  Note that the sum
of $x_1(t+1)$ and $x_2(t+1)$ determined in this way agrees with $x(t+1)$.

\iffalse
In matrix form, the preceding differential equations
for $x_1(t)$ and $x_2(t)$ can be rewritten as
\begin{equation}
\bfX'=\bfX(t)\left(\left(\begin{array}{cc}
\beta_{11}(t) p & \beta_{12}(t)(1-p) \\
\beta_{21}(t) p & \beta_{22}(t)(1-p)\end{array}\right)-\gamma(t)\bfI\right),
\label{SIR-1}
\end{equation}
where $\bfX(t)=(x_1(t), x_2(t))$.  If the parameters $\beta_{ij}(t)$ and $\gamma(t)$
are constant, the system (\ref{SIR-1}) has only one equilibrium point, which is $(0, 0)$.
This equilibrium point is stable if the eigenvalues of $\bfA-\gamma\bfI$ are all negative, where
\[
\bfA=\left(\begin{array}{cc}
\beta_{11} p & \beta_{12}(1-p) \\
\beta_{21} p & \beta_{22}(1-p)\end{array}\right).
\]
The eigenvalues of $\bfA$ are $\lambda_1$ and $\lambda_2$, where
\begin{align}
\lambda_1 &= \frac{1}{2}\left(p\beta_{11}+(1-p)\beta_{22}+\sqrt{\Delta} \right)\label{lam1}\\
\lambda_2 &= \frac{1}{2}\left(p\beta_{11}+(1-p)\beta_{22}-\sqrt{\Delta}\right),\nonumber
\end{align}
and
\begin{equation*}
\Delta =\left(p\beta_{11}-(1-p)\beta_{22}\right)^2+4(1-p)p\beta_{21}\beta_{12}.
\end{equation*}
Note that both eigenvalues are real, and $\lambda_1 > \lambda_2$.  The eigenvalues of
$\bfA-\gamma\bfI$ are $\lambda_1-\gamma$ and $\lambda_2-\gamma$, respectively.
The stability condition of the equilibrium point $(0, 0)$ is
\[
\lambda_1 < \gamma.
\]
\fi

We summarize the algorithm that determines parameters of the time-dependent SIR model in
Algorithm \ref{alg1}.
\begin{algorithm}[h]
	\caption{Estimation and prediction}\label{alg1}
	{\bf Inputs}: Sequences $\{x(t): t=1, 2, \ldots \}$, $\{r(t): t=1, 2, \ldots\}$ and parameters
	$\eta_1, \eta_2$\newline
	{\bf Outputs}: $p(t), \beta_{22}(t), x_1(t), x_2(t), S(t)$\newline
	\begin{algorithmic}[1]
		\STATE Initially at $t=1$, set $p(1)=0$, $x_1(1)=0$, $x_2(1)=y(1)$;
		\STATE Compute $\gamma(1)$ using Eq. (\ref{gamma-est});
		\STATE Compute $\beta_{22}(1)$ using Eq. (\ref{beta22-approx});
		\FOR{$t=2, 3, \ldots$}
		\STATE Find $p(t)=\hat p(t)$ by solving the optimization problem in Eq. (\ref{approx-ii})
		using parameter $\beta_{22}(t-1)$;
		\STATE Compute $\gamma(t)$ using Eq. (\ref{gamma-est});
		\STATE Find $\beta_{22}(t)$ using Eq. (\ref{beta22-approx}) and $p(t)$, $x_1(t)$,
		$x_2(t)$ and $y(t+1)$;
		\STATE Compute $x_1(t+1)$ and $x_2(t+1)$ by Eqs. (\ref{xx1}) and (\ref{xx2});
		\STATE Find predicted giant component size $S(t)$ using Eq. (\ref{S});
		\ENDFOR
	\end{algorithmic}
\end{algorithm}

\section{Maximum Likelihood Estimation of $p(t)$}\label{SEP}

In this section we present a maximum likelihood estimation
method to determine the value $p(t)$ based on $x(t)$, $x(t+1)$ and $\beta_{22}(t-1)$.
This estimation will be progressively used by Algorithm \ref{alg1} in periods $t=2, 3, \ldots$.
Thus, in the rest of this paper we simplify the notation by eliminating the dependency of time
from $p(t)$ and $\gamma(t)$ and simply use $p$ and $\gamma$, respectively.

Recall that $x(t)$ and $x(t+1)$ are the total number of infected individuals in
periods $t$ and $t+1$.   We would like to determine the value of
$p$ such that the likelihood of this sample path
is maximized.  To determine the likelihood function, we propose a
	social network model and a probabilistic version
	of the SIR model to spread the disease in this network.
	We first describe the social network model and then the probabilistic SIR model.
	It is well known that social networks possess a few important properties,
	such as significant clustering coefficients, power law degree distributions, short
	path lengths, positive degree-degree correlations, existence of
	community structures and etc.  In this paper, we propose
	a social network model that possesses significant clustering coefficients and allow
	analysis of epidemic spreading at the same time.
	Our social network model is a tree connection of cliques of random sizes arranged in layers.
	Specifically, in layer zero, there is a clique consisting of a random number of fully connected
	vertices.  Each vertex in a layer connects to $k$ cliques of random sizes in the layer.
	% In layer $i$, where $i=0, 1, \ldots$, there are $k^i$ random cliques.  Each clique consists
	% of $k_c$ fully connected vertices, and each vertex in layer $i$ connects to $k$ cliques
	% in layer $i+1$.  Thus, layer $i$ has $(k k_c)^i$ cliques and, thus has
	% $k_c(k k_c)^i$ vertices.
	We model dense clusters or households in social networks by cliques.
	Randomly connected cliques have been proposed to model social networks by Coupechoux et al. \cite{Coupechoux.2014}
	and Trapman \cite{Trapman.2007}.  We illustrate a graphical example of this model
	in \rfig{tree-cliques}.  Let $K$ be the size of a generic clique.  Assume that the probability
	mass function (pmf) of $K$ is given, i.e.
	\[
	F(i)=\pr(K=i),\qquad i=1, 2, \ldots.
	\]
	From this pmf, we compute recursively the $n$-th convolution with itself, i.e.
	\[
	F^{(n)}(i)=\sum_{j=0}^i F^{(n-1)}(i-j)F(j).
	\]
	Let $k_c$ and $\sigma_c^2$ be the mean and variance of $K$, i.e.
	\begin{align}
	&\ex[K]=k_c \label{k_c}\\
	&\mbox{Var}(K)=\sigma_c^2.\label{sigma_c}
	\end{align}
	This model clearly has a non-vanishing clustering coefficient.
	We derive the clustering coefficient of this model at the end of this section.

\bfig{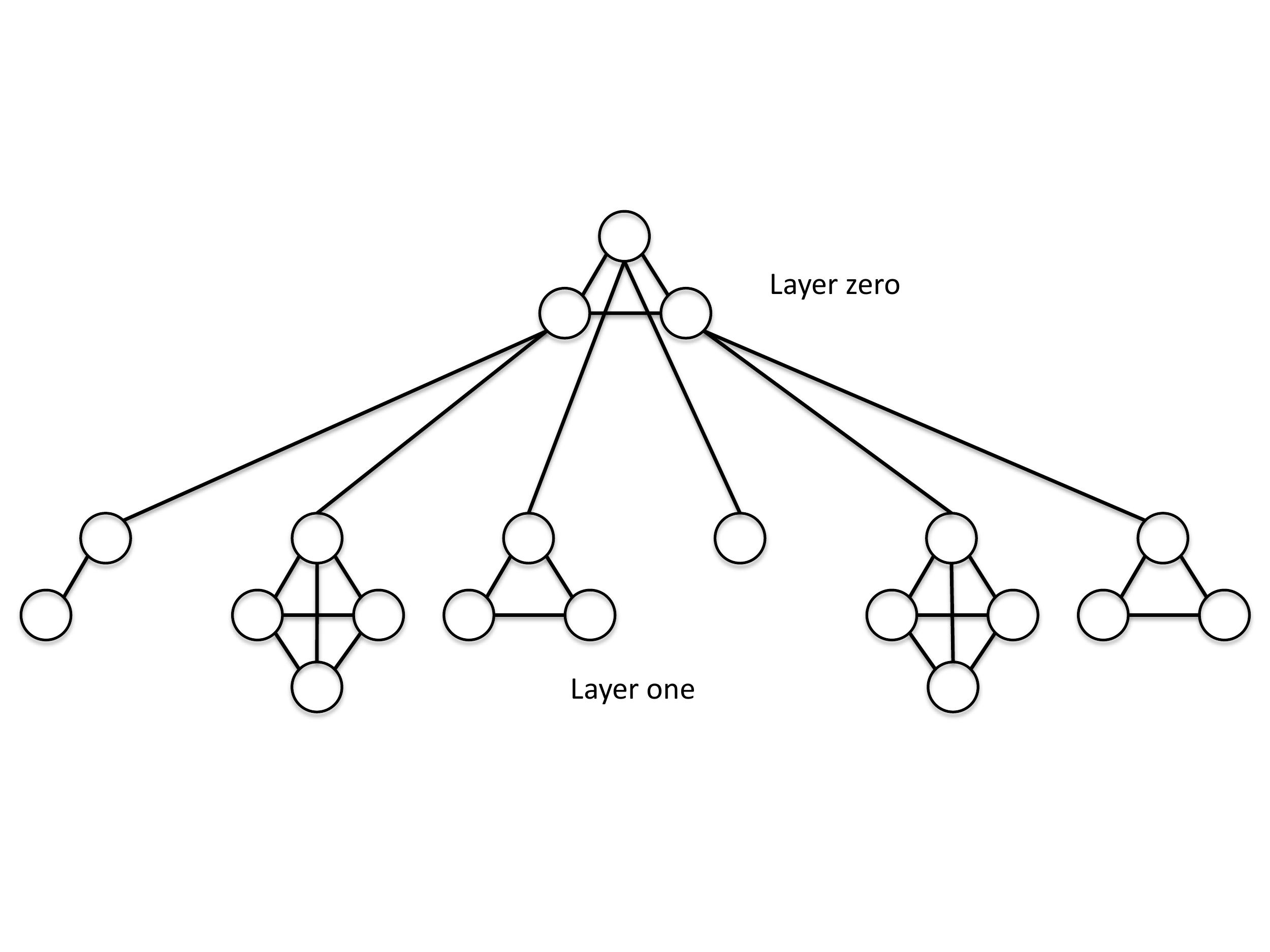}{3.5in}
\efig{tree-cliques}{An illustration of the social network, in which an epidemic spreads.
	In this example, $k=2$. The size of cliques is random.}

To determine the likelihood function, we propose a probabilistic version
of the SIR model that spreads the disease in the
social network model described in the last paragraph.  This probabilistic SIR
model is based on independent cascade models.
Independent cascade models are popular not only in the study of epidemic spreading
but also in the influence maximization problems of viral marketing \cite{Kempe2003, Easley2010}.
In an independent cascade model, each infected node has exactly one opportunity to
transmit the disease to its neighbors.  Whether the transmissions are successful or not
depend on independent events.  In our model, we distinguish between nodes that have
used their opportunity to transmit the disease from those who have not.
A node is said to be infectious, if this node has not used its opportunity to
	transmit the disease to its neighbors.  A node is said to be infected but not infectious, if
	it has used its opportunity to transmit the disease.
Our model is a discrete time model.  Let $X_t$ be the total number of currently
infected individuals in period $t$.  Let $Y_t$ be the number of individual who
contract the disease in period $t$.  In our model, we assume that a layer $i$ vertex who contracts
the disease in period $t$ have ability to transmit the disease to their neighbors in layer $i+1$
in period $t+1$, and lose the ability in periods $t+2, t+3, \ldots$ and so on.
Recall that we use cliques to model dense clustering such as
	households in a social network, Since household-based transmission is typically
	much stronger than that outside households \cite{House.2008, House.2009}, we assume that
once a vertex in a clique becomes infected, all vertices in that clique are
infected at once.   In period $t$, there are $Y_t$ infectious individuals and
$X_t-Y_t$ infected individuals who can not transmit the disease to others.
Those who are infected but cannot transmit the disease to others in period $t$
can remain infected in subsequent periods or become recovered with probability $\gamma$.
A graphical illustration of the model is shown in \rfig{bptree}.  It is clear that
this model is a discrete time Markov chain.  We shall derive the transition
probability
\begin{align}
\pr(X_{t+1}=x(t+1), Y_{t+1}=y(t+1) | X_{t}=x(t), Y_{t}=y(t))\label{transprob}
\end{align}
and find $\hat p(t)$ such that
\begin{align}
\hat p(t)=\argmax_{0\le p\le 1} \log(&\pr(X_{t+1}=x(t+1), Y_{t+1}=y(t+1) | \nonumber\\
&\quad X_{t}=x(t), Y_{t}=y(t))). \label{np}
\end{align}
We remark that our assumption that infected individuals can transmit the disease
in the beginning of their infection periods, and lose the ability to transmit in the
later part of the infection periods has nothing to do the real dynamics of the disease.
In independent cascade models each infected vertex has exactly one opportunity to transmit
the disease to its neighbors.  An infected vertex can execute its opportunity in any
one time slot during its infection period.  In this paper, we assume that infected
vertices execute their opportunities in the beginning of infection periods.
We make this assumption so that the mathematical analysis of (\ref{transprob}) is possible.
In this paper we have not attempted to model the microscopic transmission dynamics of COVID-19.
In fact, \cite{Rockx.2020, Du.2020} showed
that COVID-19 patients typically show symptoms before they can transmit the disease.

We now specify more details of the probabilistic SIR model.  Each individual
can be of type 1, or of type 2.  Each infectious individual has $k$ contacts, to whom
he or she can transmit the disease.  A type $i$ infectious individual can transmit the
disease to a type $j$ susceptible individual with probability $\phi_{ij}$, where
$i, j = 1, 2$.  Parameter $\phi_{ij}$ is related to $\beta_{22}(t)$ and $\gamma$
through
\begin{equation}\label{phi-beta}
\phi_{ij}=\frac{\beta_{ij}(t)/\gamma}{k}.
\end{equation}
Let $C$ be the number of type 1 individuals among the $Y_t$ infectious individuals
in period $t$.  Conditioning on event $\{C=i\}$,
one can express $Y_{t+1}$ in terms of $Y_t$ in the following manner.
Let
\begin{align}
I_1 &= \sum_{j=1}^{ik} \indicatorFunction{U_j=1}\\
I_2 &= \sum_{j=1}^{(Y_t-i)k}\indicatorFunction{V_j=1}.
\end{align}
$\{U_j, j=1, 2, \ldots\}$ and $\{V_j, j=1, 2, \ldots\}$ are two independent
and identically distributed (i.i.d.) sequences of Bernoulli random variables
with success probabilities $p_1$ and $p_2$, respectively.  The
two sequences are independent to anything else.
Event $\{U_j=1\}$ (resp. event $\{V_j=1\}$) indicates a type 1 (resp. type 2)
infectious individual successfully transmits
the disease to a neighboring node in the next layer.
Thus, $I_1$ (resp. $I_2$) is the number of cliques that are infected by type 1 (resp. type 2)
infectious individuals.    Recall that we assume that all members
in a clique are infected, if one member in the clique is infected.  Individuals in these cliques
are all infected and are infectious at time $t+1$.
Thus, we can express $Y_{t+1}$ in terms of $I_1$ and $I_2$, i.e.
\begin{equation}\label{inf-inf}
Y_{t+1}=\sum_{j=1}^{I_1+I_2} K_j,
\end{equation}
where we recall that $\{K_j\}$ is an i.i.d. sequence of random variables that denote
random clique sizes.  An infected individual becomes recovered with probability
$\gamma$, and remains infected otherwise.  Hence,
\beq{noninf}
X_{t+1}=Y_{t+1}+\sum_{j=1}^{X_t} \indicatorFunction{W_j=1},
\eeq
where $\{W_j, j=1, 2, \ldots\}$ is an i.i.d. sequence of Bernoulli random variable
independent of anything else.  The success probability of $W_j$ is $1-\gamma$.
Let
\beq{I3}
I_3 = \sum_{j=1}^{X_t} \indicatorFunction{W_j=1},
\eeq
and we rewrite \req{noninf} as
\beq{noninf2}
X_{t+1}=Y_{t+1}+I_3.
\eeq
We present a schematic description in \rfig{bptree}.

We now analyze $p_1$ and $p_2$, which are the success probabilities of the
Bernoulli random variables $U_j$ and $V_j$, respectively.
Event $\{U_j=1\}$ indicates a type 1 infectious individual successfully transmits
the disease to a neighboring node.  This occurs with probability $p_1$,
where \begin{equation}
p_1=\phi_{11}p+\phi_{12}(1-p) \label{def-p1}
\end{equation}
Similarly, event $\{V_j=1\}$ indicates a type 2 infectious individual successfully transmits
the disease to a neighboring node with probability $p_2$, where
\begin{align}
p_2= \phi_{21}p+\phi_{22}(1-p).\label{def-p2}
\end{align}

\bfig{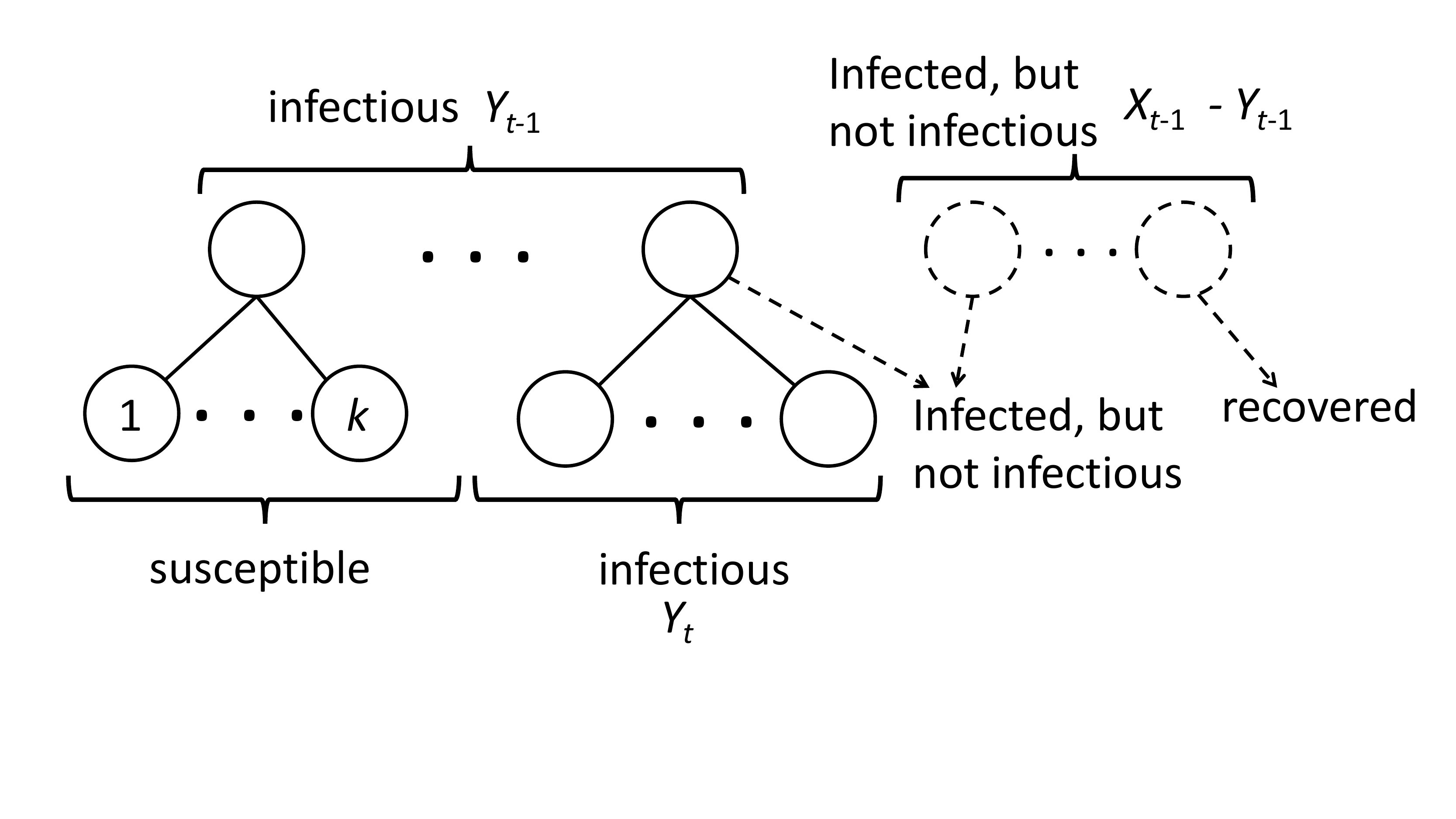}{3.5in}
\efig{bptree}{A probabilistic SIR model.  Nodes shown in solid line are infectious.
	Nodes shown in dashed line are infected, but not infectious.  Each infectious root
	node has exactly $k$ contact cliques. Infected nodes that are not
	infectious can remain infected in the next period, or become recovered.}

\iffalse
We now derive an expression for the transition probability in (\ref{transprob})
as a function of $p$.  Let
\begin{align}
I_1 &= \sum_{j=1}^{ik} \indicatorFunction{U_j=1}\\
I_2 &= \sum_{j=i+1}^{Y_t k}\indicatorFunction{V_j=1}\\
I_3 &= \sum_{j=1}^{X_t} \indicatorFunction{W_j=1}.
\end{align}
We can rewrite (\ref{inf-inf}) and \req{noninf} in terms of $I_1$, $I_2$
and $I_3$, i.e.
\begin{align*}
Y_{t+1}&= {\color{red}\sum_{j=1}^{I_1+I_2} K_j} \\
X_{t+1}&= Y_t +I_3,
\end{align*}
where $\{K_j, j=1, 2,\ldots\}$ is an independent sequence of random numbers that have
the same distribution as $K$.
\fi

Conditioning of event $\{C=i\}$ and using (\ref{inf-inf}) and \req{noninf}, we have
\begin{align}
&\pr(X_{t+1}=x(t+1), Y_{t+1}=y(t+1) | X_{t}=x(t), \nonumber\\
&\qquad Y_{t}=y(t), C=i)=\nonumber\\
&\pr(I_3=x(t+1)-y(t), \sum_{j=1}^{I_1+I_2}K_j=y(t+1) | X_{t}=x(t),\nonumber\\
&\qquad Y_{t}=y(t), C=i)\label{transprob2}
\end{align}
Conditioning on event $\{C=i, I_1=i_1, I_2=i_2, I_3=j\}$, the transition probability
is
\begin{align}
&\pr(X_{t+1}=x(t+1), Y_{t+1}=y(t+1) | X_{t}=x(t), \nonumber\\
&\qquad Y_{t}=y(t), C=i, I_1=i_1, I_2=i_2, I_3=j)\nonumber\\
&=F^{(i_1+i_2)}(y(t+1)),\label{transprob3}
\end{align}
where $F^{(n)}$ is the $n$-th convolution of the clique size pmf with itself.
Conditioning on $\{C=i\}$, random variables $I_1$, $I_2$ and $I_3$ all have
binomial distributions.
Define binomial probability mass function
\[
b(x, n, r)\defeq \left\{\begin{array}{ll}
\binomcoeff{n}{x} r^x (1-r)^{n-x} & x=0, \ldots, n \\
0 & \mbox{otherwise}. \end{array}\right.
\]
The conditional distributions of $I_1$, $I_2$ and $I_3$ are
\begin{align}
&\pr(I_1=j | X_{t}=x(t), Y_{t}=y(t), C=i)=b(j, ik, p_1)\label{conddist-I1} \\
&\pr(I_2=j | X_{t}=x(t), Y_{t}=y(t), C=i) \nonumber\\
&\quad=b(j, (y(t)-i)k, p_2) \label{conddist-I2}\\
&\pr(I_3=j | X_{t}=x(t), Y_{t}=y(t), C=i)\nonumber\\
&\quad=b(j, x(t), 1-\gamma).\label{conddist-I3}
\end{align}
Note that $I_1$, $I_2$ and $I_3$ are independent.
Unconditioning by taking average on (\ref{transprob3}) with respect to the
pmfs of $I_1, I_2$ and $I_3$ in (\ref{conddist-I1}), (\ref{conddist-I2}) and (\ref{conddist-I3}),
we obtain (\ref{transprob4}) displayed in Fig. \ref{fig-transprob4}.
\begin{figure*}[!htbp]
\begin{align}
&\pr(X_{t+1}=x(t+1), Y_{t+1}=y(t+1) | X_{t}=x(t),Y_{t}=y(t), C=i)=\nonumber\\
&\left(\sum_{i_1=0}^{ik}\sum_{i_2=0}^{(y(t)-i)k}F^{(i_1+i_2)}(y(t+1))
b(i_1, ik, p_1)b(i_2, (y(t)-i)k,p_2) \right)
 b(x(t+1)-y(t), x(t), 1-\gamma)\label{transprob4}
\end{align}
\caption{Conditional transition probability.  Variable $i$ denotes the number of type 1 infectious individuals.
	Variable $i_1$ (resp. $i_2$) denotes the number of newly infected individuals by type 1 (resp. type 2)
	infectious individuals.}\label{fig-transprob4}
\end{figure*}
In addition,
\begin{align}
\pr(C=i | Y_t=y(t))&= b(i, y(t), p).\label{dist-C}
\end{align}
Taking average on (\ref{transprob3}) with respect to event $\{C=i\}$
using (\ref{dist-C}), we have (\ref{transprob5}) displayed in Fig. \ref{fig-transprob5}.
%\begin{strip}
\begin{figure*}[!htbp]
\begin{align}
&\pr(X_{t+1}=x(t+1), Y_{t+1}=y(t+1) | X_{t}=x(t),Y_{t}=y(t))=\nonumber\\
&\sum_{i=0}^{y(t)}\sum_{i_1=0}^{ik}\sum_{i_2=0}^{(y(t)-i)k}F^{(i_1+i_2)}(y(t+1))
b(i_1, ik, p_1)b(i_2, (y(t)-i)k,p_2)
 b(x(t+1)-y(t), x(t), 1-\gamma)b(i, y(t), p)\label{transprob5}
\end{align}
\caption{Transition probability.  Variable $i$ denotes the number of type 1 infectious individuals.
	Variable $i_1$ (resp. $i_2$) denotes the number of newly infected individuals by type 1 (resp. type 2)
	infectious individuals.}\label{fig-transprob5}
\end{figure*}
%\end{strip}
In a special case, in which $K$ is degenerate and is equal to $k_c$, i.e
$\Pr(K=k_c)=1$, (\ref{transprob5}) leads to a slightly simpler
expression as follows.
%\begin{strip}
\begin{figure*}[!htbp]
	\begin{align}
	&\pr(X_{t+1}=x(t+1), Y_{t+1}=y(t+1) | X_{t}=x(t),Y_{t}=y(t))=\nonumber\\
	&\sum_{i=0}^{y(t)}\sum_{i_1=0}^{ik}\sum_{i_2=0}^{(y(t)-i)k}
	\indicatorFunction{(i_1+i_2) k_c=y(t+1)}
	b(i_1, ik, p_1)b(i_2, (y(t)-i)k,p_2)
	b(x(t+1)-y(t), x(t), 1-\gamma)b(i, y(t), p)\label{transprob-spc}
	\end{align}
	\caption{Logarithmic value of the transition probability.}\label{fig-transprob-spc}
	\end{figure*}
%\end{strip}

Both Eq. (\ref{transprob5}) and Eq. (\ref{transprob-spc}) are very complicated
to evaluate and they are difficult
to be directly used in the optimization problem (\ref{np}) when
$y(t)$ or $y(t+1)$ is large.  We now propose an
approximation method to simplify (\ref{transprob5}).
Conditioning on event $\{C=i\}$,
$I_1$ and $I_2$ are independent binomial random variables.
The mean and variance of $I_1+I_2$ are
\[
\ex[I_1+I_2 | C=i]=i k p_1+(y(t)-i)k p_2
\]
and
\[
\sigma_{I_1+I_2 | C=i}^2= ik p_1(1-p_1)+(y(t)-i)k p_2(1-p_2).
\]
Define random variable
\beq{crv}
R=\sum_{j=1}^{I_1+I_2} K_j.
\eeq
Recall that $k_c$ and $\sigma_c^2$ are the mean and
variance of a clique size, defined in (\ref{k_c})
and (\ref{sigma_c}).
The conditional mean and the conditional variance
of the compound random variable $R$ are
\begin{align}
\ex[R | C=i]&=\ex[I_1+I_2]\ex[K_1] \nonumber\\
&=(ik p_1+(y(t)-i)k p_2) k_c
\label{meanR}
\end{align}
and
\begin{align}
\mbox{Var}(R | C=i) &= \sigma_c^2 \ex[I_1+I_2 | C=i]+ (k_c)^2
\sigma_{I_1+I_2 | C=i}^2 \nonumber\\
&= (ik p_1+(y(t)-i)k p_2)\sigma_c^2 +(k_c)^2 \nonumber\\
&\qquad (ik p_1(1-p_1)+(y(t)-i)k p_2 (1-p_2))\label{varR}
\end{align}
according to Wald's equations \cite[p. 438--439]{Ghahramani2005}.
Since $y(t)$ is typically very large, $I_1$ and $I_2$ are also large.
We propose to approximate $R$ by a normal random variable.  That is,
\begin{align}
&\pr(R\in(y(t+1)-1/2, y(t+1)+1/2) | C=i) \nonumber\\
&=\frac{\exp\left(\frac{-(y(t+1)-\ex[R|C=i])^2}{2\mbox{\scriptsize Var}(R|C=i)}\right)}
{\sqrt{2\pi\mbox{\small Var}(R|C=i)}}.\label{normal-approx}
\end{align}
We further simplify (\ref{normal-approx}) by applying mean-field approximation
\cite{Newman2010, Wipf2012}.  We replace $C$ by its mean value, which is
$y(t)p$.  Replacing $i$ by $y(t)p$ in (\ref{meanR}) and (\ref{varR}) and then
assuming that $p_1$ and $p_2$ are small, we reach
\begin{align}
\mu &=\ex[R] \nonumber\\
&= y(t)(p p_1+(1-p) p_2)k k_c \nonumber\\
&=y(t) R_0 \label{mu}
\end{align}
and
\begin{align*}
\sigma^2 &=\mbox{Var}(R) \nonumber\\
&= \sigma_c^2 y(t) k (p p_1+(1-p) p_2)+(k_c)^2 y(t) k \nonumber\\
&\qquad \cdot (pp_1(1-p_1)+(1-p)p_2(1-p_2))\nonumber\\
&\approx\left(\frac{\sigma_c^2}{k_c}+k_c\right)y(t)R_0,
\end{align*}
where
\beq{defR0}
R_0=(p p_1+(1-p) p_2)k k_c
\eeq
is the basic reproduction number of this epidemic network.  Let
\[
\xi=\frac{\sigma_c^2}{k_c}+k_c
\]
and rewrite
\begin{equation}\label{sigma}
\sigma^2=\xi y(t)R_0.
\end{equation}
Taking logarithm on (\ref{transprob5}) and noting (\ref{normal-approx}), we obtain
(\ref{logtransprob}). Eq. (\ref{logtransprob}) is displayed in Fig. \ref{fig-logtransprob}.
%\begin{strip}
\begin{figure*}[!htbp]
	\begin{align}
	&\log\pr(X_{t+1}=x(t+1), Y_{t+1}=y(t+1) | X_{t}=x(t),
	Y_{t}=y(t))\nonumber\\
	&\approx \log\left(\frac{\exp\left(\frac{-(y(t+1)-\mu)^2}{2\sigma^2}\right)}
	{\sqrt{2\pi\sigma^2}}\right)+\log(b(x(t+1)-y(t), x(t), 1-\gamma))\nonumber\\
	&=\log\left(\frac{\exp\left(\frac{-(y(t+1)-y(t)R_0)^2}{2\xi y(t)R_0}\right)}
	{\sqrt{2\pi\xi y(t)R_0}}\right)+\log(b(x(t+1)-y(t), x(t), 1-\gamma))
	\label{logtransprob}
	\end{align}
	\caption{Approximation of transition probabilities by normal distributions.}
	\label{fig-logtransprob}
	\end{figure*}
%\end{strip}
Since the second term on the right side of (\ref{logtransprob}) is independent of
$p$, maximizing the logarithmic transition probability is equivalent to maximizing the
first term on the right side.
Define
\begin{align}
f(R_0) &\defeq -\frac{(y(t+1)-y(t) R_0)^2}{2\xi y(t) R_0}-\frac{1}{2}\log(2\pi\xi y(t))\nonumber\\
&\quad-\frac{1}{2}\log(R_0).\label{approx-2}
\end{align}
We approximate the optimization problem in (\ref{np}) by the following optimization problem
\begin{equation}\label{approx-ii}
\hat p(t)=\argmax_{0\le p\le 1} f(R_0)
\end{equation}
with constraint $R_0\ge 0$.
To study the extrema of $f$ we differentiate $f$ once and twice to get
\begin{align}
f'(R_0) &= \frac{y(t+1)^2-y(t)^2 R_0^2 -\xi y(t)R_0}{2\xi y(t)R_0^2} \label{f'}\\
f''(R_0) &= \frac{\xi y(t)R_0-2y(t+1)^2}{2\xi y(t) R_0^3}.\label{f"}
\end{align}
Note that
\beq{f'=0}
f'(R_0)=0
\eeq
has a unique positive root, which is
\beq{root-of-f'}
R_0=\frac{-\xi+\sqrt{\xi^2+4 y(t+1)^2}}{2y(t)}.
\eeq
At this root, the second derivative of $f$ is negative.  To see this, note that the positive root
in \req{root-of-f'} satisfies
\[
\xi y(t)R_0 = y(t+1)^2-y(t)^2 R_0^2.
\]
Substituting the preceding into (\ref{f"}), we obtain
\begin{align*}
f''(R_0) &= \frac{\xi y(t)R_0-2y(t+1)^2}{2\xi y(t) R_0^3} \\
&= \frac{-y(t)^2 R_0^2-y(t+1)^2}{2\xi y(t) R_0^3} \\
&<0.
\end{align*}
Thus,
\[
\log\pr(X_{t+1}=x(t+1), Y_{t+1}=y(t+1) | X_{t}=x(t),
Y_{t}=y(t))
\]
achieves maximum when \req{root-of-f'} holds.  In view of \req{defR0}, $R_0$ is
a quadratic function of $p$. We solve $p$ from \req{root-of-f'}.
Recall that we assume (\ref{beta12-assumption}), (\ref{beta21-assumption})
and (\ref{beta11-assumption}).  Under these assumptions,
\req{root-of-f'} can have a unique root of $p$ in $[0, 1]$, in which
case, the unique root is the solution of the optimization problem
(\ref{approx-ii}).  Eq. \req{root-of-f'} can have no root of $p$
in $[0, 1]$, in which case the solution of (\ref{approx-ii}) is
$\hat p(t)=0$ or $\hat p(t)=1$.  We summarize the
solution of the optimization problem (\ref{approx-ii}) in the following
proposition.  The proof of the proposition is presented in the appendix
at the end of this paper.
\bprop{result-approx-ii}
Eq. \req{root-of-f'} has two real roots in $p$.  Either \req{root-of-f'} has
exactly one root in interval $[0, 1]$, or it has no root in this interval.
If \req{root-of-f'} has exactly one root in $[0, 1]$, it is the smaller
root (denoted by $p^\star$) of the two roots. In this case, the optimal
solution of (\ref{approx-ii}) is $p^\star$.  If Eq. \req{root-of-f'} has no
roots in $[0, 1]$, either $p^\star < 0$ or $p^\star > 1$.  In the former case,
the optimal solution is $\hat p(t)=0$.  In the latter case, the optimal solution
is $\hat p(t)=1$.
\eprop

\subsection{Clustering Coefficient}\label{CC}

In this section we analyze the clustering coefficient of the social network model
	described in section \ref{SEP}. According to Newman \cite[Eq. (7.41)]{Newman2000},
	the clustering coefficient of a network is defined as
	\beq{def-CC}
	M=\frac{3\ex[T]}{\ex[D]},
	\eeq
	where $\ex[T]$ is the expected number of triangles, and
	$\ex[D]$ is the expected number of connected
	triples in the network.  To derive these two quantities, let $\{K_{ij}\}$
	be an i.i.d.  double sequence of random variables that has the same distribution
	of $K$.  Random variable
	$\{K_{ij}\}$ denotes the size of the $j$-th clique in layer $i$.
	Let $N_i$ (resp. $C_i$) be the number of vertices (resp. cliques) in layer $i$.
	It is easy to see that $N_i$ and $C_i$ satisfy the following recursion
	\begin{align}
	N_i &= \sum_{j=1}^{C_{i}} K_{ij} \label{recur1}\\
	C_{i+1} &= k N_i,\nonumber
	\end{align}
	starting from $C_0=1$ and $N_0=K_{01}$.
	$N_i$ defined in (\ref{recur1}) is a compound random variable, whose
	mean can be evaluated using Wald's equations \cite{Ghahramani2005}.  That is,
	\begin{align*}
	\ex[N_i] &= \ex[C_{i}]\ex[K]=k_c \ex[C_{i}]\\
	\ex[C_{i+1}] &= k \ex[N_{i}].
	\end{align*}
	It can be shown by easy induction that the preceding two equations lead
	to the following
	\begin{align}
	\ex[N_i] &= k_c (kk_c)^i \label{exN}\\
	\ex[C_i] &= (k k_c)^i.\label{exC}
	\end{align}
	
	We now derive the expected number of triangles and the expected number of
	connected triples in the social network described in section \ref{SEP}.
	Suppose that the network has $L$ layers from indexed from  zero to $L-1$.
	The number of triangles in layer $i$ is
	\beq{Ti}
	T_i=\sum_{j=1}^{N_i}\binom{K_{ij}}{3}\indicatorFunction{K_{ij}\ge 3},
	\eeq
	$i=0, 1, \ldots, L-1$.
	Since
	\[
	\ex[K^j | K\ge 3]=\frac{\sum_{x=3}^\infty x^j \Pr(K= x)}{\Pr(K\ge 3)}
	\]
	for any $j\ge 1$, one can express $\ex[T_i]$ more compactly in terms
	of the conditional expectation $\ex[K^j | K\ge 3]$.  Specifically,
	\begin{align}
	\ex[T_i] &= \ex[N_i] \Pr(K\ge 3)\cdot \nonumber\\
	&\quad \frac{\ex[K^3 | K\ge 3] -3\ex[K^2 | K\ge 3] +2\ex[K | K\ge 3]}{6}.\label{eTi}
	\end{align}
	
There are four types of connected triples.  An example to illustrate the triangles and connected triples
is shown in \rfig{cc}. In the first type of connected triples, both edges
connect vertices in a clique.  The expected number of type 1 connected triples in layer $i$ is
\[
D_{1,i}=\sum_{j=1}^{N_i}K_{ij}\binom{K_{ij}-1}{2}\indicatorFunction{K_{ij}\ge 3} ,\quad i=0, 1, \ldots, L-1.
\]
It follows from the same argument leading to (\ref{eTi}) that
\begin{align*}
\ex[D_{1,i}] &=\ex[N_i] \Pr(K\ge 3)\cdot \nonumber\\
&\quad \frac{\ex[K^3 | K\ge 3] -3\ex[K^2 | K\ge 3] +2\ex[K | K\ge 3]}{2}.
\end{align*}	
We have
\beq{n21}
3 \ex[T_i]=\ex[	D_{1,i}].
\eeq
In the second type of connected triples, one edge connects two vertices in a layer
$i$ clique, and the second
edge connects the two vertices with a vertex in layer $i+1$.  The number of type 2
connected triples is
\[
D_{2,i}=\sum_{j=1}^{N_i} K_{ij} (K_{ij}-1)k,
\ i=0, 1, \ldots, L-2.
\]
It follows that
\beq{n22}
\ex[D_{2,i}]=\ex[N_i] (\ex[K^2]-\ex[K])k.
\eeq
A third type of connected triple connects a vertex in layer $i$ with two vertices in layer
$i-1$ and $i+1$.  The number of type 3 connected triples is
\[
D_{3,i}=k N_i,\quad i=1, 2, \ldots, L-2.
\]
It follows that
\beq{n23}
\ex[D_{3,i}]=k \ex[N_i].
\eeq
In the fourth type of connected triples, one edge connects two vertices in a layer
$i$ clique, and the second
edge connects the two vertices with a vertex in layer $i-1$.  The number of type 4
connected triples is
\[
D_{4,i}=\sum_{j=1}^{N_i} (K_{ij}-1),\quad i=1, 2, \ldots, L-1.
\]	
It follows that
\beq{n24}
\ex[D_{4,i}]=\ex[N_i] (\ex[K]-1).
\eeq
Substituting (\ref{eTi}), \req{n21}, \req{n22}, \req{n23} and \req{n24}
into \req{def-CC}, we have (\ref{M}) displayed in Fig. \ref{fig-M}.

%\begin{strip}
\begin{figure*}[!htbp]
\begin{align}
M&=\frac{3\sum_{\ell=0}^{L-1} \ex[T_i]}{3\sum_{\ell=0}^{L-1}\ex[T_{i}]+\sum_{i=0}^{L-2} \ex[D_{2,i}]
	+\sum_{i=1}^{L-2} \ex[D_{3,i}]+\sum_{i=1}^{L-1}\ex[D_{4,i}]}.\label{M}
\end{align}
\caption{Clustering coefficient.}\label{fig-M}
\end{figure*}
%\end{strip}
	In a special case, in which $K$ is degenerate and is equal to $k_c$, one can further simplify
	(\ref{M}).  Particularly, letting $L\to\infty$, one obtains
	\[
	M\to \left\{\begin{array}{ll}
	\frac{k_c^2(k_c-1)(k_c-2)}{k_c^2(k_c-1)(k_c-2)+4k_c(k_c-1)+2} & k_c\ge 3\\
	0 & k_c=1, 2.\end{array}\right.
	\]

\bfig{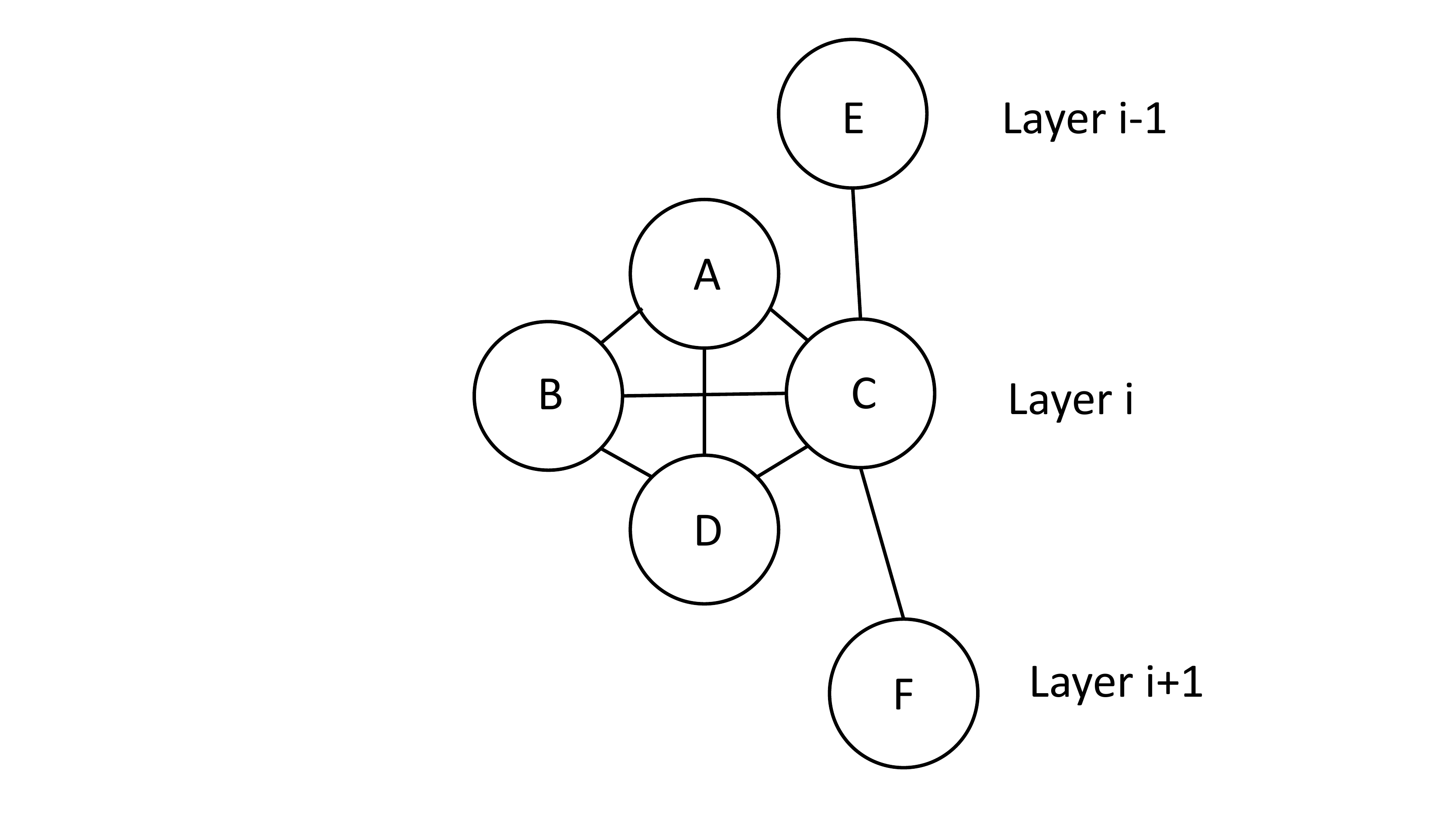}{3.5in}
\efig{cc}{An illustration of triangles and connected triples.  In this example, $k_c=4$ and $k$
	is not shown.  Vertices $A, B, C$ and $D$ form a clique in layer $i$.   Vertices $A, B$ and $C$
	form a triangle.  $BAC$ is a type one connected triple.  $DCF$ is a type two connected triple.
	$ECF$ is a type three connected triple and $ECD$ is a type four connected triple.}

\section{Percolation Analysis}\label{SPA}

In this section we consider a random contact network, in which a disease transmitted by droplets
or aerosols spreads according to an independent cascade model.  That is, the disease transmits
from a node at one end of an edge to the node at the other end with a probability.  In addition,
the transmissions along all edges are independent.  We shall present a percolation
analysis of this model and obtain percolation thresholds and sizes of giant components.

We now describe our model.  Consider a random graph $(G, V, E)$.
Randomly select a node from the graph.  Let $Z$ be the degree of this node.
Let $g_0(z)$ denote the probability generating function of $Z$, i.e.,
\begin{equation}\label{def-eta0}
g_0(z)=\sum_{k=0}^\infty \pr(Z=k)z^k.
\end{equation}
Now randomly select an edge.
Let $Y$ be the excess degree of a node reached along the
randomly selected edge.  Let $g_1(z)$ be the probability generating function
of $Y$, {\em i.e.}
\[
g_1(z)=\sum_{k=0}^\infty \pr(Y=k) z^k.
\]
Every node in this graph can be one of two types.  A type 1 node denotes an individual who
wears a facial mask and a type 2 node denotes an individual who does not.
A randomly selected node is of type 1 with probability $p$ and is of type 2 with probability
$1-p$.  Assume that this event is independent of anything else.
As mentioned before, an infectious disease spreads in this network according to
an independent cascade model.  Consider a randomly selected edge connecting two nodes,
say node $V_1$ and node $V_2$.
Let $\phi_{ij}$ be the conditional probability that the disease transmits from
node $V_1$ to node $V_2$, given that the types of nodes $V_1$ and $V_2$
are $i$ and $j$, respectively, where $i, j=1, 2$.

We now present a percolation analysis of the random network model described
above.  Percolation analysis has been a useful tool to study resilience
of communication networks \cite{Cohen2000, Hu2011} and epidemic networks
\cite{Sander2002, Moore2000, Cardy1985, serrano2006percolation}.
There are two types of percolation models, bond percolation and site percolation,
depending on whether edges or nodes, respectively, are removed randomly
\cite{Newman2010}.  Our model is a form of bond percolation, where edges
are randomly removed.  Removed edges imply that diseases can not  be transmitted
from one end of these edges to the other end.  The size of the largest component
in the percolated network is the maximum fraction of possibly infected population
in the network.

Randomly select an edge. Let $V_1$ and $V_2$ be the two nodes at the two ends
of the edge.  Suppose that the type of node $V_1$ is $i$.
Let $E_{i}$ be the event that along the selected edge from $V_1$,
one {\em can not} reach a giant component.  Let $u_{i}=\pr(E_{i})$.  Now we
condition on the event that the type of node $V_1$ is $j$.
Event $E_{i}$ will occur, if the randomly selected edge is removed.  This occurs with
probability $1-\phi_{ij}$.  With probability $\phi_{ij}$ the randomly selected edge is
present.  Let $Y$ be the number of neighbors of node $V_2$, not including $V_1$.
Event $E_{i}$ will occur, if one can not reach a giant component along
any one these $Y$ edges.  Combining these arguments, we have
\begin{align}
u_{i}&=p\left(1-\phi_{i1}+\phi_{i1}\sum_{k=0}^\infty \pr(E_{1} | Y=k)\pr(Y=k)\right)\nonumber\\
&\quad+(1-p)\left(1-\phi_{i2}+\phi_{i2}\sum_{k=0}^\infty \pr(E_{2} | Y=k)\pr(Y=k)\right)\nonumber\\
&=p\Biggl(1-\phi_{i1}+\phi_{i1}\sum_{k=0}^\infty \sum_{\ell=0}^k \pr(E_{1} |
Y=k, C=\ell)\nonumber\\
&\qquad \times\pr(C=\ell | Y=k)\pr(Y=k)\Biggr)\nonumber\\
&\quad+(1-p)\Biggl(1-\phi_{i2}+\phi_{i2}\sum_{k=0}^\infty \sum_{\ell=0}^k \pr(E_{2} |
Y=k, C=\ell)\nonumber\\
&\qquad \times\pr(C=\ell | Y=k)\pr(Y=k)\Biggr),\label{u_ij}
\end{align}
where $C$ is the number of type 1 nodes connected to node $V_2$ not including $V_1$.
The distribution of $C$ conditioning on $Y=k$ is binomial.
Thus,  Eq. (\ref{u_ij}) becomes
\begin{align}
u_{i}&=p\left(1-\phi_{i1}+\phi_{i1}\sum_{k=0}^\infty \sum_{\ell=0}^k
(u_{1}^\ell\cdot u_{2}^{k-\ell})\right.\nonumber\\
&\qquad\times\binomcoeff{k}{\ell}p^\ell (1-p)^{k-\ell}
\pr(Y=k)\Biggr) \nonumber\\
&\quad+(1-p)\left(1-\phi_{i2}+\phi_{i2}\sum_{k=0}^\infty \sum_{\ell=0}^k
(u_{1}^\ell\cdot u_{2}^{k-\ell})\right.\nonumber\\
&\qquad\times\binomcoeff{k}{\ell}p^\ell (1-p)^{k-\ell}
\pr(Y=k)\Biggr) \nonumber\\
&=p(1-\phi_{i1}+\phi_{i1} g_1(p u_{1}+(1-p)u_{2}))\nonumber\\
&\quad+(1-p)(1-\phi_{i2}+\phi_{i2} g_1(p u_{1}+(1-p)u_{2}))\nonumber\\
&=1-(p\phi_{i1}+(1-p)\phi_{i 2})(1-g_1(p u_{1}+(1-p)u_{2}))\label{u}
\end{align}
for $i=1, 2$. Eq. (\ref{u}) is a system of nonlinear equations, from which we can solve
for $u_1$ and $u_2$.

Once we have $u_{1}$ and $u_2$ we can compute the giant component size of the percolated network.
Randomly select a node from the network.
Let $S_i$, where $i=1, 2$, be the conditional probability that the randomly selected
node is connected with a giant component, given that the selected node is of type $i$.
Let $X$ be the degree of the randomly selected node.
The node is connected with a giant component if along at least one
of its edges one can reach a giant component.  Conditioning on $X=k$, let $I$ be the
number of type 1 nodes among the $k$ neighbors.  Combining all these arguments, we have
\begin{align*}
S_i&=1-\sum_{k=0}^\infty\sum_{j=0}^k u_{1}^j u_{2}^{k-j}\pr(X=k)
\binomcoeff{k}{j}p^j (1-p)^{k-j}\\
&= 1-g_0(p u_{1}+(1-p)u_{2})
\end{align*}
for $i=1, 2$.
By taking average on the conditional probabilities, a randomly selected node is connected
to a giant component with probability
\begin{equation}
S=p S_1+(1-p)S_2.\label{S}
\end{equation}
This is also the expected size of the giant component.

\subsection{Percolation Threshold}

Let $\bfu$ be a $2\times 1$ vector over the set of real numbers $\cal R$.
Specifically, let
\[
\bfu=(u_{1}, u_{2})^T,
\]
where symbol $T$ denotes transposition of vectors.  Let $\bff$ be a vector-valued
function mapping from ${\cal R}^2$ to ${\cal R}^2$, {\em i.e.,}
\begin{align}
&\bff(\bfx) = (f_1(\bfx), f_2(\bfx))^T = \nonumber\\
&\left(\begin{array}{l}
1-(p\phi_{11}+(1-p)\phi_{11})(1-g_1(px_{1}+(1-p)x_{2})) \\
1-(p\phi_{21}+(1-p)\phi_{22})(1-g_1(px_{1}+(1-p)x_{2}))
\end{array}\right),\label{def-f}
\end{align}
where $\bfx=(x_1, x_2)$.
Eq. (\ref{u}) implies that $\bfu$ is a root of
\begin{equation}\label{fixpoint1}
\bfx=\bff(\bfx).
\end{equation}
The roots of equations of the form (\ref{fixpoint1}) are also called the
fixed points of function $\bff$.  It is clear that function $\bff$ always
has fixed point $\bfone=(1, 1)^T$.
Lee {\em et al.} \cite{Lee2019} established
that $\bff$ has an additional fixed point if the dominant eigenvalue of
the Jacobian matrix evaluated at $\bfone$ is greater than one.
In addition, this fixed point is attractive.
The Jacobian matrix for function $\bff$ evaluated at $\bfx=\bfa$ is defined as
\begin{equation}\label{Jacobian}
\left.\left(\begin{array}{cc}
\frac{\partial f_1(\bfsx)}{\partial x_1} & \frac{\partial f_1(\bfsx)}{\partial x_2}  \\
\frac{\partial f_2(\bfsx)}{\partial x_1} & \frac{\partial f_2(\bfsx)}{\partial x_2}
\end{array}\right)\right|_{\bfsx=\mathbf{a}}
\end{equation}
The Jacobian matrix for the function $\bff$ defined in (\ref{def-f}) evaluated
at $(1,1)^T$ is
\begin{equation}
\bfJ=\ex[Y]\left(
\begin{array}{cc}
\phi_{11}p & \phi_{12}(1-p) \\
\phi_{21}p & \phi_{22}(1-p)\end{array}\right),
\label{J1}
\end{equation}
where $\ex[Y]=g_1'(1)$ is the expected excess degree of a node reached by a randomly
selected edge.  It is easy to derive the eigenvalues of $\bfJ$.
Denote the two eigenvalues of $\bfJ$ by $\lambda_1$ and $\lambda_2$, Then,
\begin{align}
\lambda_1 &= \frac{\ex[Y]}{2}\left(p\phi_{11}+(1-p)\phi_{22}+\sqrt{\Delta} \right)\label{lam1}\\
\lambda_2 &= \frac{\ex[Y]}{2}\left(p\phi_{11}+(1-p)\phi_{22}-\sqrt{\Delta}\right),\nonumber
\end{align}
where
\begin{equation*}
\Delta =\left(p\phi_{11}-(1-p)\phi_{22}\right)^2+4(1-p)p\phi_{21}\phi_{12}.
\end{equation*}
Note that both eigenvalues are real, and $\lambda_1 > \lambda_2$.
Thus, spectral radius or the dominant eigenvalue of $\bfJ$ is $\lambda_1$.
It is also this eigenvalue that controls the percolation threshold of the epidemic
network.

\section{Numerical and Simulation Results}\label{SNS}

In this section we present numerical and simulation results.
We first verify the accuracy of (\ref{logtransprob}) to approximate the
transition probability in
(\ref{transprob4}).  We select an arbitrary set of parameters:
\[
k=3, x_0=5, \phi_{22}=0.8.
\]
In this section, we assume that $K$ is deterministic and $K=k_c$.
According to \cite{Brienen2010}, the efficiency of a typical mask is
in the range from 0.58\% to 85\%.  Since masks are typically more efficient
to stop viral transmission if sources wear masks \cite{Patel2016Respiratory},
we set $\eta_1=0.8$ and $\eta_2=0.65$.
We select a value for $p$ between 0 and 1.  We randomly generate ten thousand values
for $y(t)$.  We perform simulation to generate ten thousand values for $y(t+1)$
based on $y(t)$ and $p$.
We solve (\ref{np}) based on (\ref{transprob-spc}) and (\ref{approx-ii}) for
ten thousand values of $y(t)$ and $y(t+1)$.
We calculate the average distance between the two solutions for $k_c=1$
and $k_c=3$.  The result is shown in \rfig{accuracy}.  This shows that the
approximation method works very well for $k_c=1$.  The approximation error for
$k_c=3$ is also acceptable.

\iffalse
\bfigx{distance1.eps}{3.5in}
\efig{accuracy}{Distance between the solution of the constraint
	optimization using exact transition probability in (\ref{transprob4}) and the approximated
	transition probability in (\ref{approx-ii}).}
\fi

\bfig{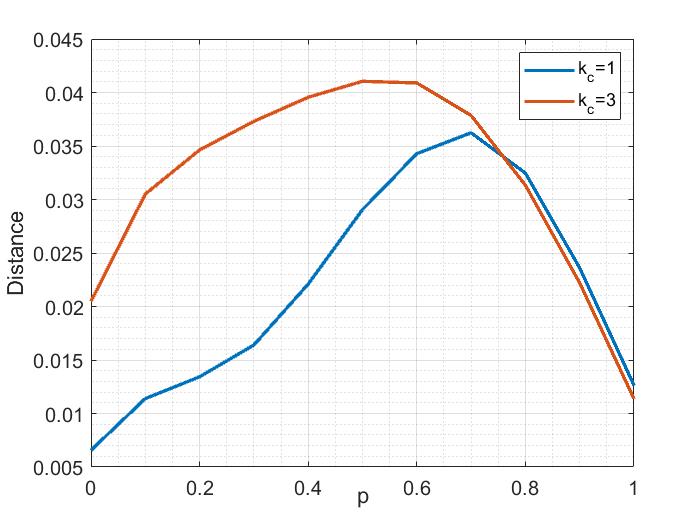}{3.5in}
\efig{accuracy}{Distance between the solution of the constraint
	optimization using exact transition probability in (\ref{transprob-spc}) and the approximated
	transition probability in (\ref{approx-ii}).}

Next, we use the data published by John Hopkins University \cite{John-Hopkins2020} to
predict the spread of the COVID-19 pandemic using our model.  In this study, we assume
that $K$ is deterministic and $K=k_c=3$.
Due to large infection rates of some countries, we assume that
$k=\ex[Y]=150$ in order to have $\phi_{22}\le 1$.
We choose $\tau=5$ days as the width of a time slot in the time dependent SIR model.
We first execute Algorithm \ref{alg1} on the number of infected and recovered individuals
in mainland China.  The time functions of $\beta_{22}(t)$ and $p(t)$ are shown
in \rfig{China}.  The maximum fraction of population who will be infected predicted
at time $t$ is also shown in the figure.   The epidemic started from the People's Republic of
China from January, 2020.  The authority managed to contain the epidemic very well
by the end of February.  However, there are several significant events between January and August
of 2020.  In mid April, the Chinese authority revised the way that death toll is calculated
\cite{China-event-1}.  In mid June, Beijing faced a second wave of infections \cite{China-event-2}
In mid July, there was a surge of infected number of individuals in Xinjiang \cite{China-event-3}.
At the end of July, there was a surge in Dalian \cite{China-event-4}.  From \rfig{China}, we see that
$\beta_{22}(t)$ is large initially in January, and is controlled to reach a small
value by the end of February.  The value of $\beta_{22}(t)$ again rises and drops before
and after the events described above.  From \rfig{China}, note that $p(t)$ also rises
and decreases before and after the events.  However, function $p(t)$ lags behind function
$\beta_{22}(t)$.  This correlation has a nice interpretation.  Rise in $\beta_{22}(t)$
usually manifests itself in the rise of infected number of individuals.
As the population sees a rise in the infected number of individuals, more people wear
masks to protect themselves.  On the other hand, decreases of $\beta_{22}(t)$ result in
less number of newly infected individuals.  People typically see this as sign of
a safe community.  Thus, less people wear masks in public places.  As $\beta_{22}(t)$ rises,
the predicted size of giant component $S(t)$ also increases.  However, as $p(t)$
catches up and rises, $S(t)$ decreases.  Notice that the time sequence $\beta_{22}(t)$
reflects a joint efforts of many measures to contain the epidemic.
Wearing facial masks is a measure in the personal
level.  Shutting down schools, businesses, and keeping people at home is a measure in
the government level.  In this paper, we consider only the measure of wearing masks and ignore
other measures.  Thus, $p(t)$ in \rfig{China} may be higher than the actual fraction
of population who wear masks, as it reflects a joint effect of many measures to contain the
epidemic.   Note that the value of $k_c$ can influence the value of
	$p(t)$, $\beta_{22}(t)$ and the predicted size of giant components $S(t)$.  In \rfig{kc}
we show the time function $p(t)$ for $k_c=3$ and $k_c=1$.  The more densely clustered people
are, the larger $k_c$ is and the more infectious the disease is.  From \rfig{kc}, we see that
population react and more people wear masks if $k_c$ is large.

\iffalse
\begin{table}[h]
	\begin{center}
		\begin{tabular}{|l|r|r|r|r|r|} \hline
			Country	& $\phi_{22}$  & $p$ & $R_0$
			& $\lambda_1$ & $S$ \\ \hline\hline
			U.S.A. & 0.213 & 0.2980 & 1.3080 & 1.540 & 0.4865 \\ \hline
			China (mainland) & 0.875 & 0.7990 & 1.5174 & 2.2481
& 0.5139
\\ \hline
			United Kingdom & 0.189 & 0.3525 & 1.0461 & 1.2704 & 0.2959 \\ \hline
			France   & 0.258 & 0.3973 & 1.3055 & 1.6267 & 0.4856 \\ \hline
			Spain & 0.163 & 0.3231 &  0.9548 & 1.1402 & 0.1807 \\ \hline
			Italy   & 0.079 & 0.1696 & 0.6075 & 0.6654 & 0 \\ \hline
			Iran  & 0.125 & 0.1425 &  0.0049 & 1.0843 & 0.1357 \\ \hline
			South Korea   & 0.044  & 0  & 0.4399  & 0.44  & 0 \\ \hline	
	\end{tabular}\end{center}
	\caption{Values of $\phi_{22}$,
		$p$, $\lambda_1$, $R_0$
		and $S$ for various counties. \label{countries}}
\end{table}
\fi

\iffalse
\bfigx{mainlandchina.eps}{3.5in}
\efig{China}{Time functions $\beta_{22}(t)$, $p(t)$, $S(t)$ of mainland China.}
\fi

\bfig{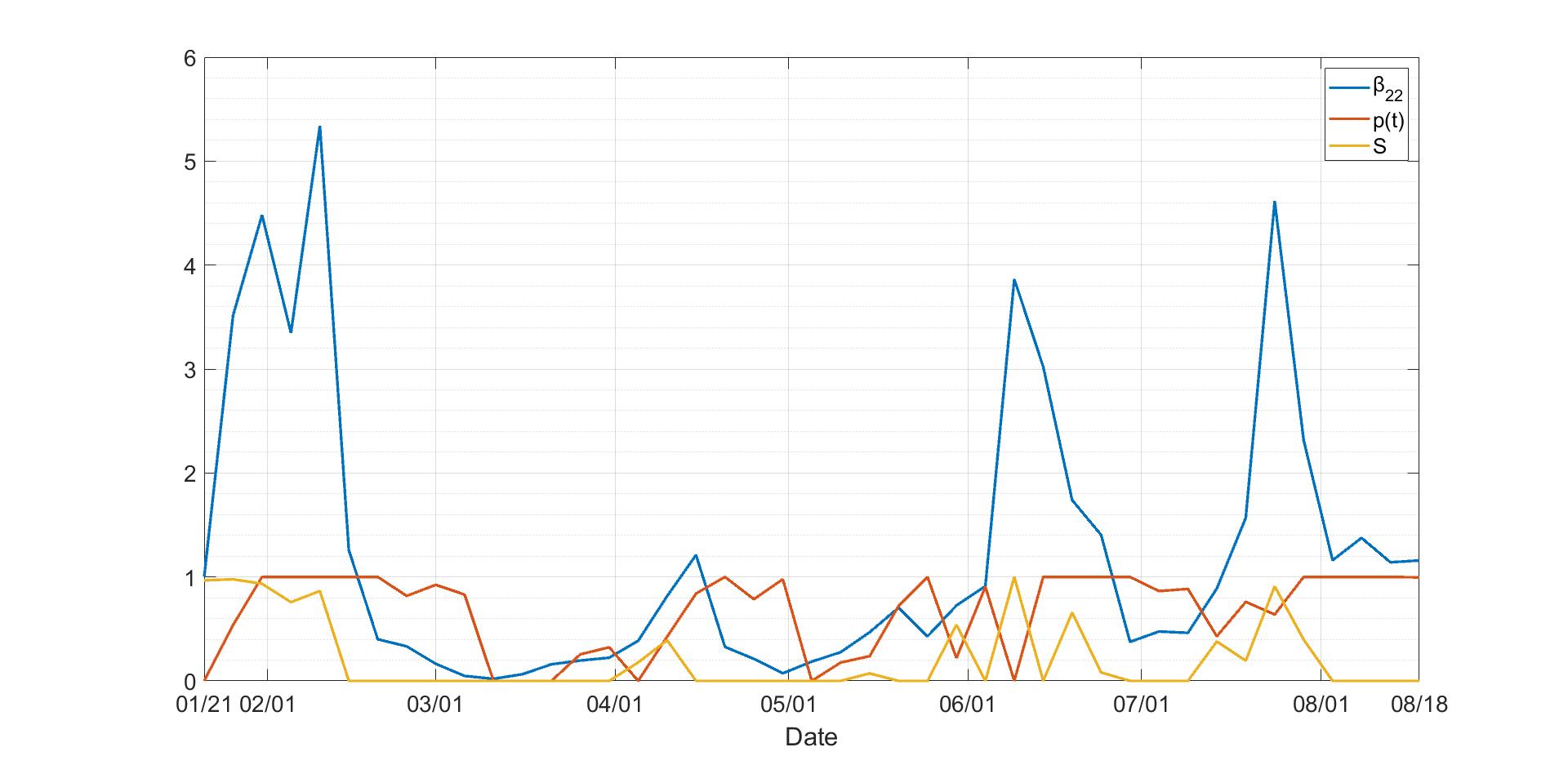}{3.5in}
\efig{China}{Time functions $\beta_{22}(t)$, $p(t)$, $S(t)$ of mainland China.}

\bfig{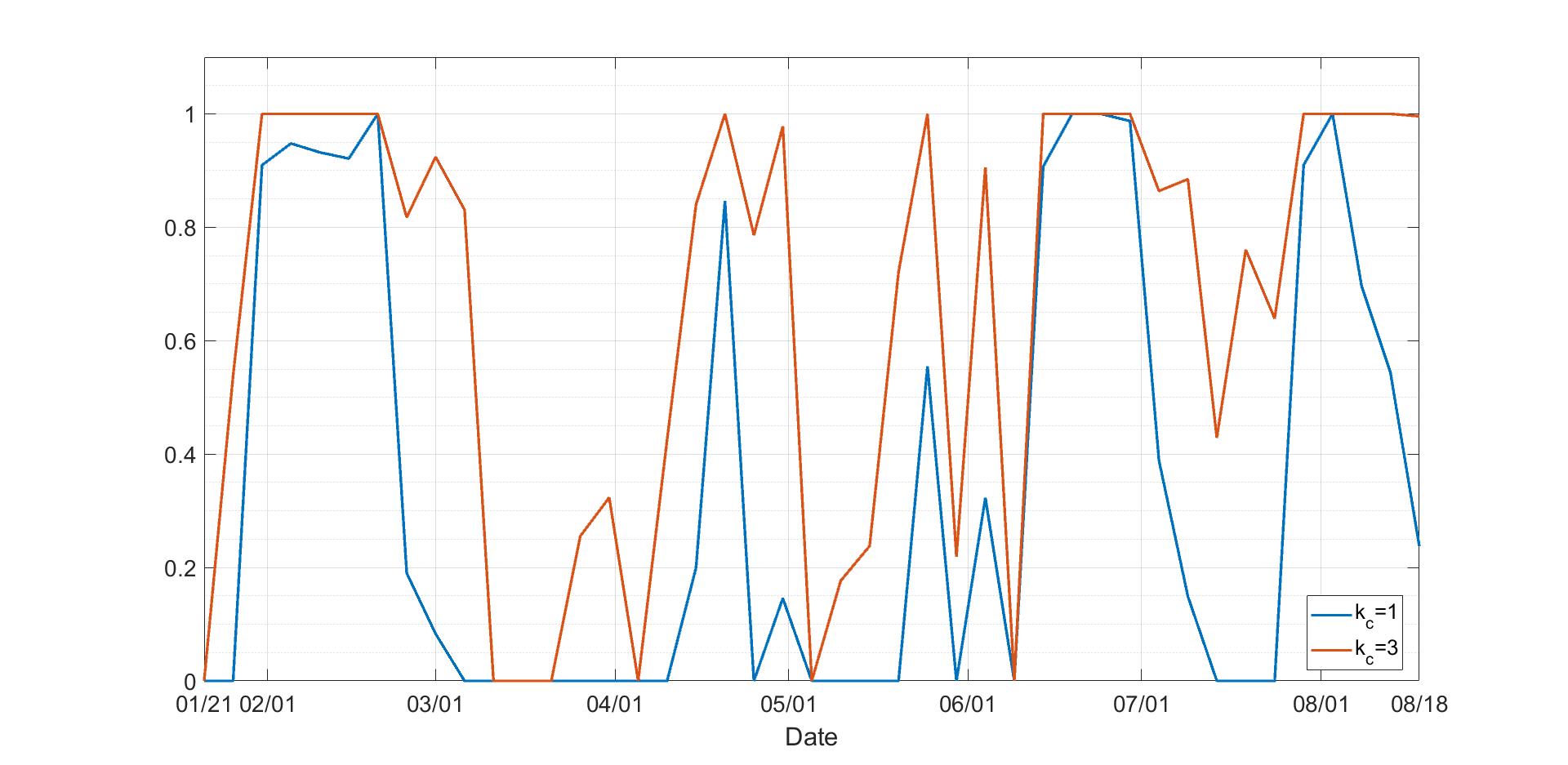}{3.5in}
\efig{kc}{$p(t)$ of mainland China from Jan. 21st to Aug. 18, 2020
	for $k_{c}=1$ and $k_c=3$.}

Next we study the epidemics of six countries.  They are the United States of America,
India, France, Italy, South Korea and Japan.
We execute Algorithm \ref{alg1} to compute $\beta_{22}(t)$, $p(t)$ and $S(t)$.
The time functions of $\beta_{22}(t)$, $p(t)$ and $S(t)$ for the six countries
are shown in \rfig{beta22}, \rfig{p} and \rfig{S}, respectively.
We present $p(t)$ and $S(t)$ of the six
 countries in \rfig{p} and \rfig{S}.
Since these functions fluctuate a lot, we use a built-in Matlab function
to compute the polynomial regression of these functions in order to show a general trend.
Note that India has a smaller infection rate $\beta_{22}(t)$ than that of the
U.S.A.  However, India has a much smaller predicted size of giant component
than that of the U.S.A.  This is because India has a very large recovery rate
\cite{India-1, India-2}.  We also note that around July France has a $\beta_{22}(t)$
comparable with that of the US.  However, France has a high predicted $S(t)$ after August.
It is worth noting that $p(t)$ of France in mid August is not close to 1.
By raising $p(t)$, France may have a better control of the epidemics.
Finally, we show the average value of $p(t)$ for three western countries versus
that of three Asian countries.  The average value of $p(t)$ of western countries
and Asian countries rises nearly equally quickly in the beginning of March 2020.
Asian countries had a better control of the epidemic.  As a result, Asian people relaxed
and $p(t)$ was significantly lower in April and May 2020. The rise and drop of $p(t)$ for
Japan and South Korea in \rfig{p} and \rfig{ave} during April and May 2020
seem to agree with the timing of
some major outbreaks recorded in \cite{Korea-event,Japan-event}.

Finally, we compare the size of giant components obtained in (\ref{S}) with that obtained
by simulation of a few well-known data sets \cite{snapnets}.  The result is shown in Table \ref{tbl1}.
In this study, we set $p=0.4$ and $\phi_{22}=0.6$. In each simulation, we randomly select
a vertex and then randomly select two vertices that are distance $i$ from the vertex for
$i=1, 2, \ldots, 5$.  These eleven individuals are infected at time zero.  We simulate the
independent cascade model until no new infection is observed.  We repeat the simulation fifty times
before we take an average.  From Table \ref{tbl1}, we note that simulation results of real-life
networks are consistently smaller than those obtained from (\ref{S}).  Note also that real-life
social networks nearly always possess clustering and community structures, while the random network
model that leads to (\ref{S}) does not.    Dense connections within a community are helpful to the
spreading of the disease within the community \cite{House.2008, House.2009}.
However, community boundaries can hinder the spreading of the disease.

\iffalse
\bfigx{total.jpg}{3.75in}
\efig{beta22}{$\beta_{22}(t)$ of the U.S., India and France.}
\bfigx{regression-p.jpg}{3.5in}
\efig{p}{$p(t)$ of the U.S., India and France.}
\bfigx{regression-S.jpg}{3.5in}
\efig{S}{Predicted size of giant component of the U.S., India and France.}
\fi

\bfig{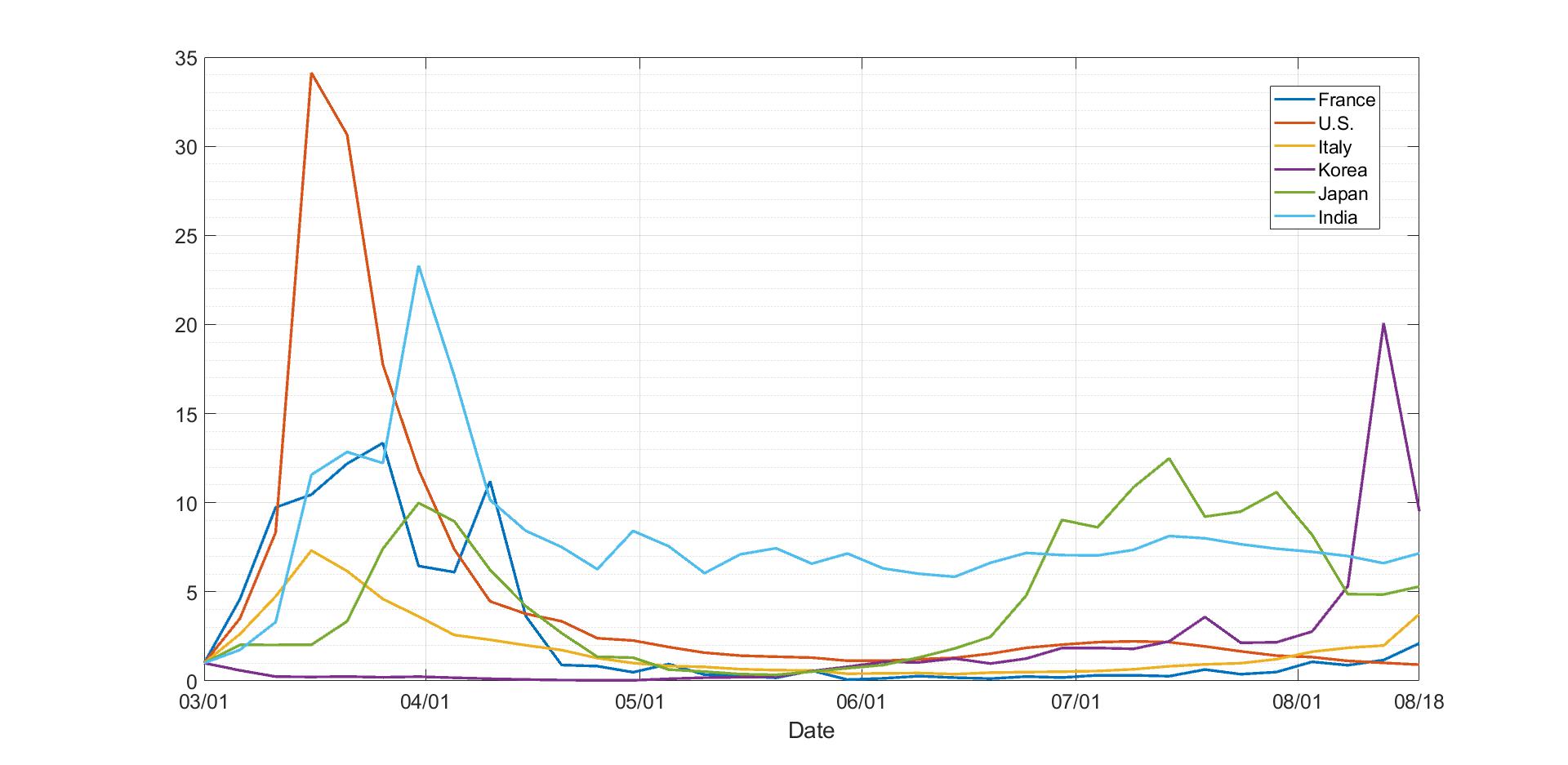}{3.75in}
\efig{beta22}{$\beta_{22}(t)$ of the U.S., India, France, Italy, Korea and Japan.}
\bfig{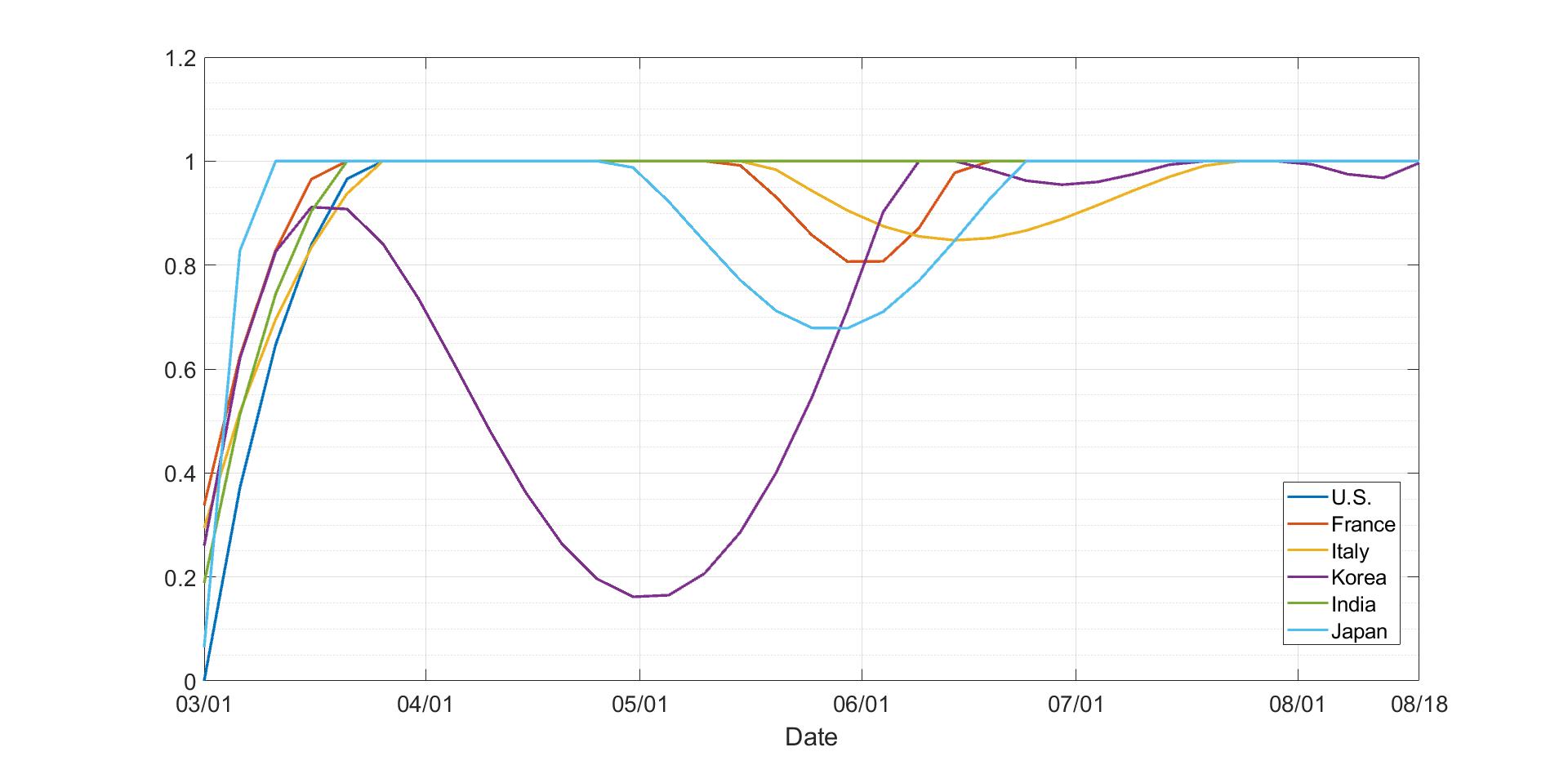}{3.5in}
\efig{p}{$p(t)$ of the U.S., India, France, Italy, Korea and Japan.}
\bfig{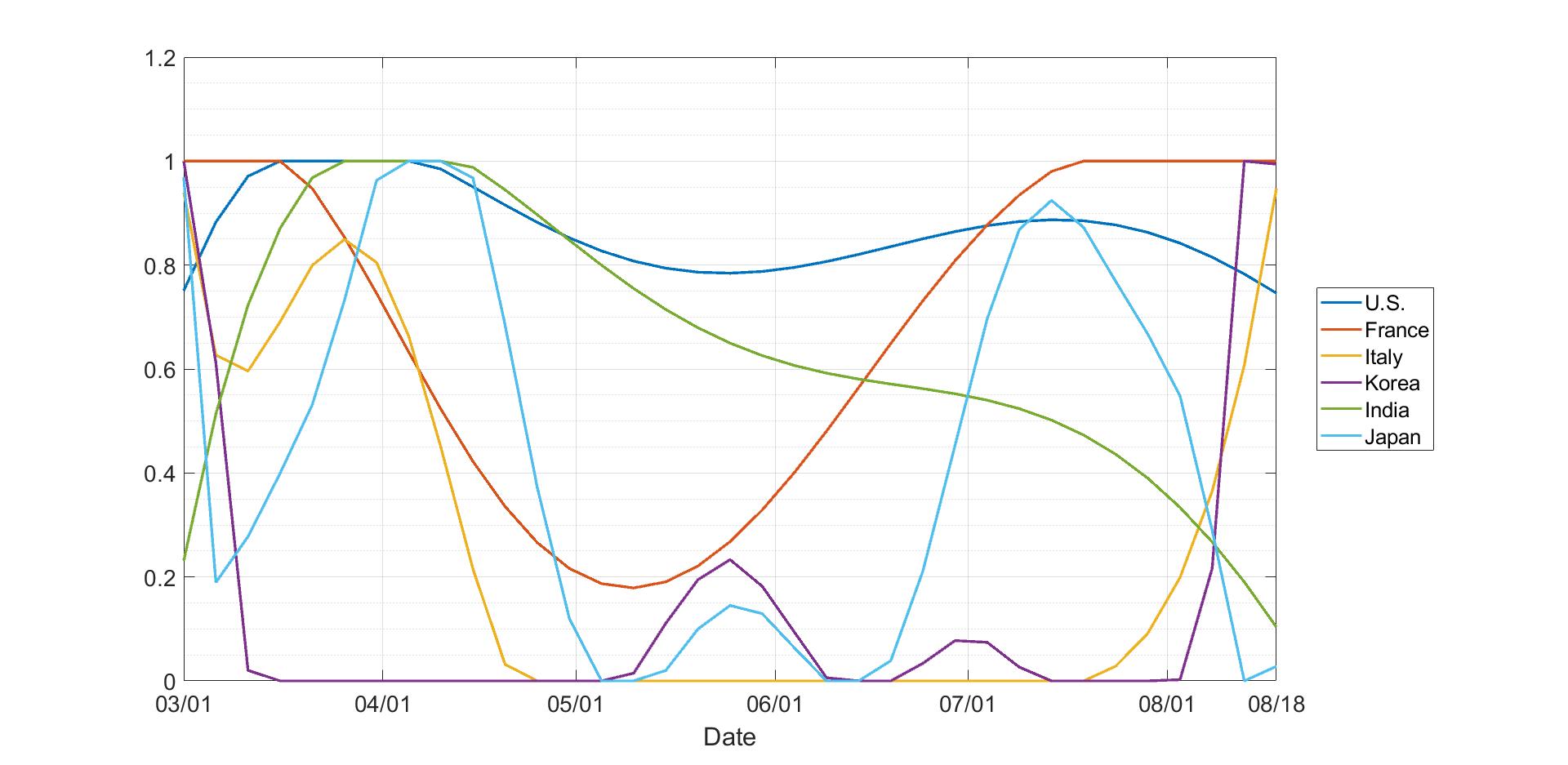}{3.5in}
\efig{S}{Predicted size of giant component of the U.S., India, France, Italy, Korea and Japan.}

\bfig{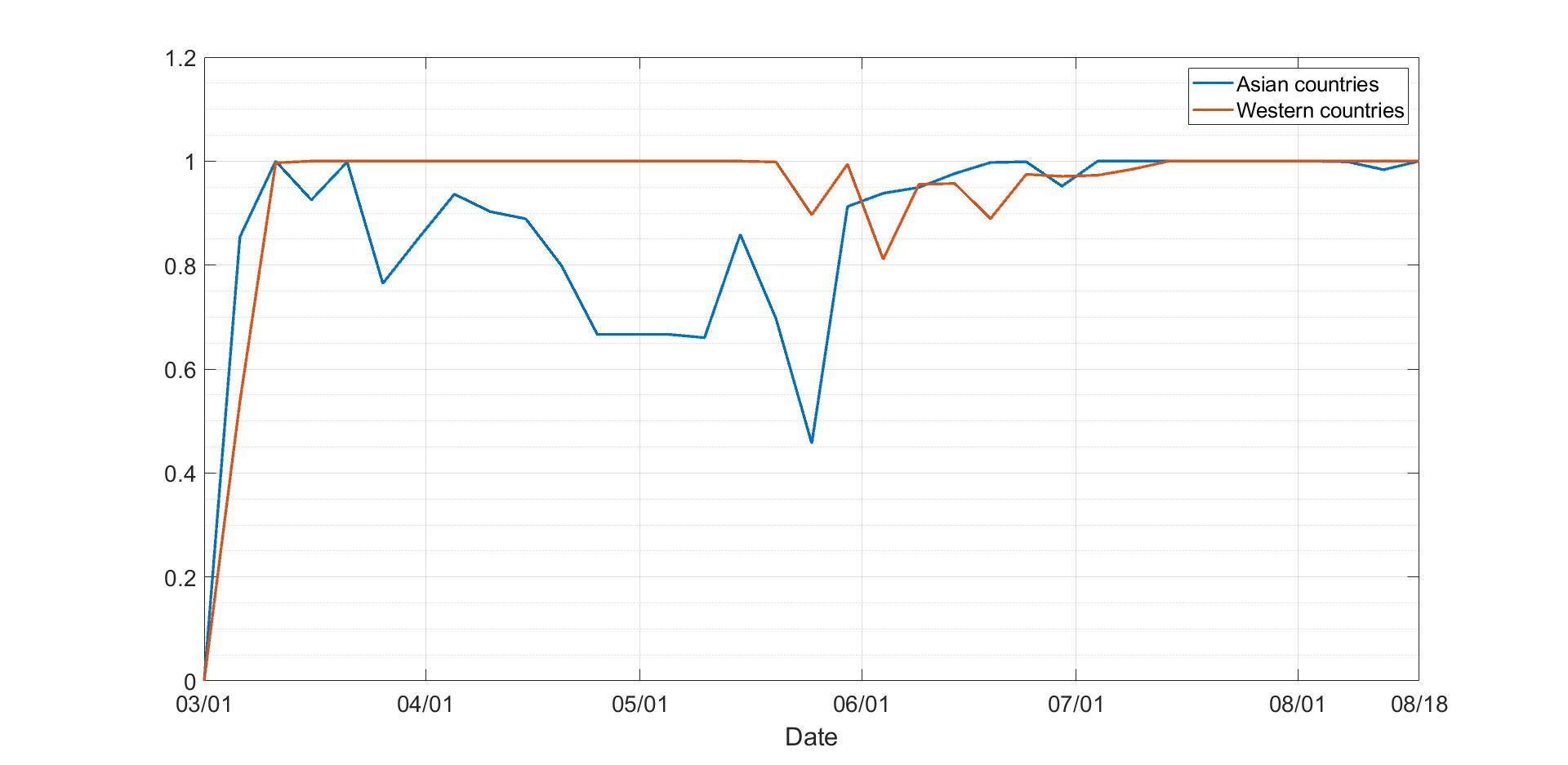}{3.5in}
\efig{ave}{Average value of $p(t)$ of three western counties versus that of three Asian countries.}

\begin{table}\begin{center}
	\begin{tabular}{|l|r|r|r|r|}\hline
		network & number of & number of & $r(\infty)$ & $S$ \\
		& vertices & edges & & \\ \hline\hline
		email-Eu-core & 1005 & 25571 & 0.782 & 0.862 \\ \hline
		Facebook & 4039 & 88234 & 0.744 & 0.898 \\ \hline
		twitch-ES & 4648 & 59382 & 0.724 & 0.803 \\ \hline
		twitch-EN & 7126 & 35324 & 0.571 & 0.640 \\ \hline
		lastfm asia & 7624 & 27806 & 0.490 & 0.577 \\ \hline
		ca-CondMat & 23133 & 93497 & 0.535 & 0.645 \\ \hline
		deezer\_europe & 28281 & 92752 & 0.500 & 0.579 \\ \hline	
	\end{tabular}\end{center}\label{tbl1}
\caption{SIR spreading in eight real-life networks.  Symbol $r(\infty)$ denotes the fraction of
vertices that are ever infected by the disease in simulation.  $S$ is the corresponding
value obtained by calculating Eq. (\ref{S}).}
\end{table}

\section{Conclusions}\label{SC}

In this paper, we presented a time dependent SIR model, in which
some individuals wear facial masks and some do not.  Based on the number
of infected individuals and the number of recovered individuals published
by J. H. University, we estimate the disease infection rates and recovery
rates.  We proposed a probabilistic version of the SIR model.
We derived the transition probability of this random SIR model.
By maximizing the transition probability, we estimate the most probable
value of the fraction of population who wear masks.
This transition probability numerically difficult to compute, if the
states of model are large.  Based on central limit theorem, we proposed
an approximation.  Through numerical and simulation study, we show that
the approximation works well.   Finally we carried out a percolation analysis
to predict the eventual fraction of population who will be infected with the disease.
We proposed a progressive analysis of the epidemics.  Using results from the
progressive analysis, we analyzed the epidemics of six countries.

\medskip
\centerline{\large  Appendix}

In this appendix, we prove  \rprop{result-approx-ii}.

{\bf Proof of \rprop{result-approx-ii}.} We rewrite $R_0$ defined in
\req{defR0} to show dependency with $p$ explicitly, i.e.
\begin{align}
R_0(p)&=k k_c \phi_{22}(\eta_1\eta_2 p^2-(\eta_1+\eta_2)p +1).
\label{defR0p}
\end{align}
In (\ref{defR0p}) we have assumed (\ref{beta12-assumption}),
(\ref{beta21-assumption}) and (\ref{beta11-assumption}).
Note that $\eta_1 < 1$ and $\eta_2 < 1$.
Since the right side of (\ref{defR0p}) is a quadratic polynomial
in $p$, it is easy to establish the following results.
\begin{enumerate}
	\item $R_0(0) > R_0(1) > 0$.
	\item $R_0'(0)=kk_c\phi_{22}(-\eta_1-\eta_2)<0$.
	\item $
	R_0'(1)=kk_c\phi_{22}(\eta_1(\eta_2 -1)+\eta_2(\eta_1 -1))<0.$
	\item The minimum of $R_0(p)$ occurs at
	\[
	p=\frac{1}{2\eta_1}+\frac{1}{2\eta_2}.
	\]
	This point is greater than one, since $\eta_1 < 1$ and $\eta_2 < 1$.
	\item The minimum of $R_0(p)$ is
	\[
	-kk_c\phi_{22}\frac{\eta_1\eta_2(\eta_1 -\eta_2)^2}{4\eta_1^2 \eta_2^2},
	\]
	which is negative.
\end{enumerate}
From the results above and the fact that the right side of \req{root-of-f'}
is positive, it follows that \req{root-of-f'} must have two real
roots.  It also follows from the results above that $R_0(p)$ is monotonically
decreasing and \req{root-of-f'} can not have two roots in $[0, 1]$.
Since $R_0(p)$ is decreasing for $p\in[0, 1]$, it follows that if
\req{root-of-f'} has exactly one root in $[0, 1]$, the root is
$p^\star$.  It is also quite clear that if \req{root-of-f'} has a
unique root in $[0, 1]$, the root is the optimal solution of
(\ref{approx-ii}).

Now we analyze cases, in which \req{root-of-f'} does not have roots
in $[0, 1]$.  Since $R_0(p)$ is decreasing for $p\in[0, 1]$, it follows that
either both roots of \req{root-of-f'} are greater than 1, or
$p^\star < 0$ and the other root is greater than 1.  We now analyze
these two cases separately.  We first note that we have treated $R_0$
as an independent variable in (\ref{approx-ii}).  Thus, the
derivative $f'(R_0)$ is with respect to $R_0$.  In the two cases in
which Eq. \req{root-of-f'} has no roots in $[0, 1]$, we have to
treat the objective function in (\ref{approx-ii}) as a function
of $p$.  Specifically, the derivative of the objective function
with respect to $p$ is
\begin{equation}
\frac{d}{dp} f(R_0(p))=f'(R_0)R_0'(p), \label{f'2}
\end{equation}
where $f'(R_0)$ is given in (\ref{f'}).
We claim that in the first case the objective function
of (\ref{approx-ii}) is increasing for all $p\in [0, 1]$.
Thus, the optimal solution of (\ref{approx-ii}) is $\hat p(t)=1$.
We also claim that in the second case, the objective function
of (\ref{approx-ii}) is decreasing for all $p\in [0, 1]$.
Thus, the optimal solution of (\ref{approx-ii}) is $\hat p(t)=0$.

We now prove the two claims.  From (\ref{f'}) and (\ref{f'2}),
\begin{align}
\frac{d}{dp} f(R_0(p))&=\frac{h(R_0)R_0'(p)}{2\xi y(t) R_0^2},\label{f'3}
\end{align}
where
\[
h(R_0)=y(t+1)^2-y(t)^2 R_0^2 -\xi y(t)R_0.
\]
Note that $h$ is a quadratic and concave function of $R_0$ with
$h(0)=y(t+1)^2>0$.  It is clear that
equation $h(R_0)=0$ has two real roots.
Function $h$
has a zero at $R_0=c$, where $c$ is the right side of \req{root-of-f'}, i.e.
\[
c=\frac{-\xi+\sqrt{\xi^2+4 y(t+1)^2}}{2y(t)}.
% c=\frac{-n+\sqrt{n^2+4 {\color{red}(y(t+1)/k_c)}^2}}{2y(t)}.
\]
If $R_0(1) < c < R_0(0)$, equation $h(R_0)=0$ has exactly one root in the closed
interval $[0, 1]$.  This root is $R_0=c$.
If $c < R_0(1)$, both roots of \req{root-of-f'} are greater than 1.
This corresponds to the first case.  If $c > R_0(0)$, then $p^\star < 0$
and the other root is greater than 1.  This corresponds to the second case.
In the first case,
\[
R_0(p) > c
\]
for all $p\in [0, 1]$.  Thus,
\[
h(R_0(p)) < h(c)=0
\]
for all $p\in [0, 1]$.
From (\ref{f'3}), it follows that $d f(R_0(p))/dp > 0$, since
$R_0'(p)<0$ for all $p\in[0, 1]$.  This proves the first claim.
In the second case,
\[
R_0(p) < c
\]
for all $p\in [0, 1]$.  Thus,
\[
h(R_0(p)) > h(c)=0
\]
for all $p\in [0, 1]$. From (\ref{f'3}), it follows that $d f(R_0(p))/dp < 0$.
This proves the second claim.

\iffalse
\centerline{\bf Acknowledgment}

This research was supported in part by the Ministry of Science and Technology,
Taiwan, R.O.C., under Contract 109-2221-E-007-093-MY2.
\fi

%\bibliography{epidemics-dbase}
%\bibliography{/Users/Duan-ShinLee/Dropbox/latex/bibdatabase.bib}
%\bibliography{/Users/lds/Dropbox/latex/bibdatabase.bib}
\bibliography{bibdatabase.bib}
\bibliographystyle{ieeetran}

\end{document}